\documentclass{article}

     \usepackage[preprint]{neurips_2019}

\usepackage[utf8]{inputenc} 
\usepackage[T1]{fontenc}    
\usepackage{hyperref}       
\usepackage{url}            
\usepackage{booktabs}       
\usepackage{amsfonts}       
\usepackage{amsmath}
\usepackage{nicefrac}       
\usepackage{microtype}      
\usepackage{graphicx}
\usepackage{subcaption}
\usepackage{wrapfig}
\usepackage[ruled,vlined]{algorithm2e}
\usepackage{algorithmic}
\usepackage{enumitem}
\usepackage{amsthm}
\usepackage{todonotes}

\newtheorem{theorem}{Theorem}

\SetKwInput{KwInput}{In}
\SetKwInput{KwOutput}{Out}

\title{Safe Interactive Model-Based Learning}

\author{%
  Marco Gallieri\thanks{Corresponding author.}
  \And
  Seyed Sina Mirrazavi Salehian
  \And
  Nihat Engin Toklu 
  \And
  Alessio Quaglino
  \And 
  Jonathan Masci
  \And
  Jan Koutn\'{i}k
    \And
  Faustino Gomez  \AND \vspace{-0.2cm} \\
  NNAISENSE, 
  Lugano, Switzerland - Austin, Texas\\  \smallskip
  \texttt{\{marco,sina,engin,alessio,jon,jan,tino\}@nnaisense.com} \\ 
}

\begin{document}

\maketitle

\begin{abstract}
Control applications present hard operational constraints. A violation of these can result in \emph{unsafe} behaviour. 
This paper introduces {Safe} Interactive Model Based Learning (SiMBL),
a framework to refine an existing controller and a system model
while operating on the real environment.
SiMBL is composed of the following trainable components:
a Lyapunov function, which determines a safe set; a safe control policy; and a Bayesian RNN forward model. A min-max control framework, based on alternate minimisation and backpropagation through the forward model, is used for the offline computation of the controller and the safe set.
Safety is formally verified {\em a-posteriori} with a probabilistic method that utilizes the Noise Contrastive Priors (NPC) idea to build a Bayesian RNN forward model with an additive state  uncertainty estimate which is large outside the training data distribution. 
%
Iterative refinement of the model and the safe set is achieved thanks to a novel 
 loss that conditions the uncertainty estimates 
of the new model to be close to the current one. 
The learned safe set and model can also be used for safe exploration, i.e., 
to collect data within the safe invariant set, for which a simple one-step MPC is proposed. 
The single components are tested on the simulation of an inverted pendulum with limited torque and stability region, showing that  iteratively adding more data can improve the model, the controller and the size of the safe region.  
\end{abstract}


\section{Approach rationale}

Safe Interactive Model-Based Learning (SiMBL) aims to control a deterministic dynamical system:
\begin{equation} \label{eq:ode}
    x(t+1) = x(t)+ dt\ f(x(t), u(t)),\quad y(t) = x(t),
\end{equation}
where $x$ is the state and $y$ are the measurements, in this case assumed equivalent. The system (\ref{eq:ode}) is sampled with a known constant time $dt$ and it is subject to closed and bounded, possibly non-convex, operational constraints on the state and input:
\begin{eqnarray}
x(t)\in\mathbb{X}\subseteq \mathbb{R}^{n_x}, \ 
u(t)\in\mathbb{U}\subset \mathbb{R}^{n_u}, \quad \forall t>0. \label{eq:constraints}
\end{eqnarray}
The stability of (\ref{eq:ode}) is studied using discrete time systems analysis. In particular, tools from discrete-time \emph{control Lyapunov functions} \citep{Blanchini, Khalil_book} will be used to compute policies that can keep the system \emph{safe}.   
 
\paragraph{Safety.} 
In this work, \emph{safety} is defined as the capability of a system to remain within a subset $\mathbb{X}_s \subseteq \mathbb{X}$ of the operational constraints and to return asymptotically to the equilibrium state from anywhere in $\mathbb{X}_s$. A feedback control policy, $u=K(x)\in\mathbb{U}$, is certified as \emph{safe} if it can provide safety with \emph{high probability}. In this work, safety is verified with a statistical method that extends  \cite{bobiti_samplingdriven_nodate}. 

\paragraph{Safe learning.} \begin{wrapfigure}{r}{7.5cm}
\centering
       \includegraphics[width=0.9\linewidth, trim={0. 0. 0. 0.}, clip]{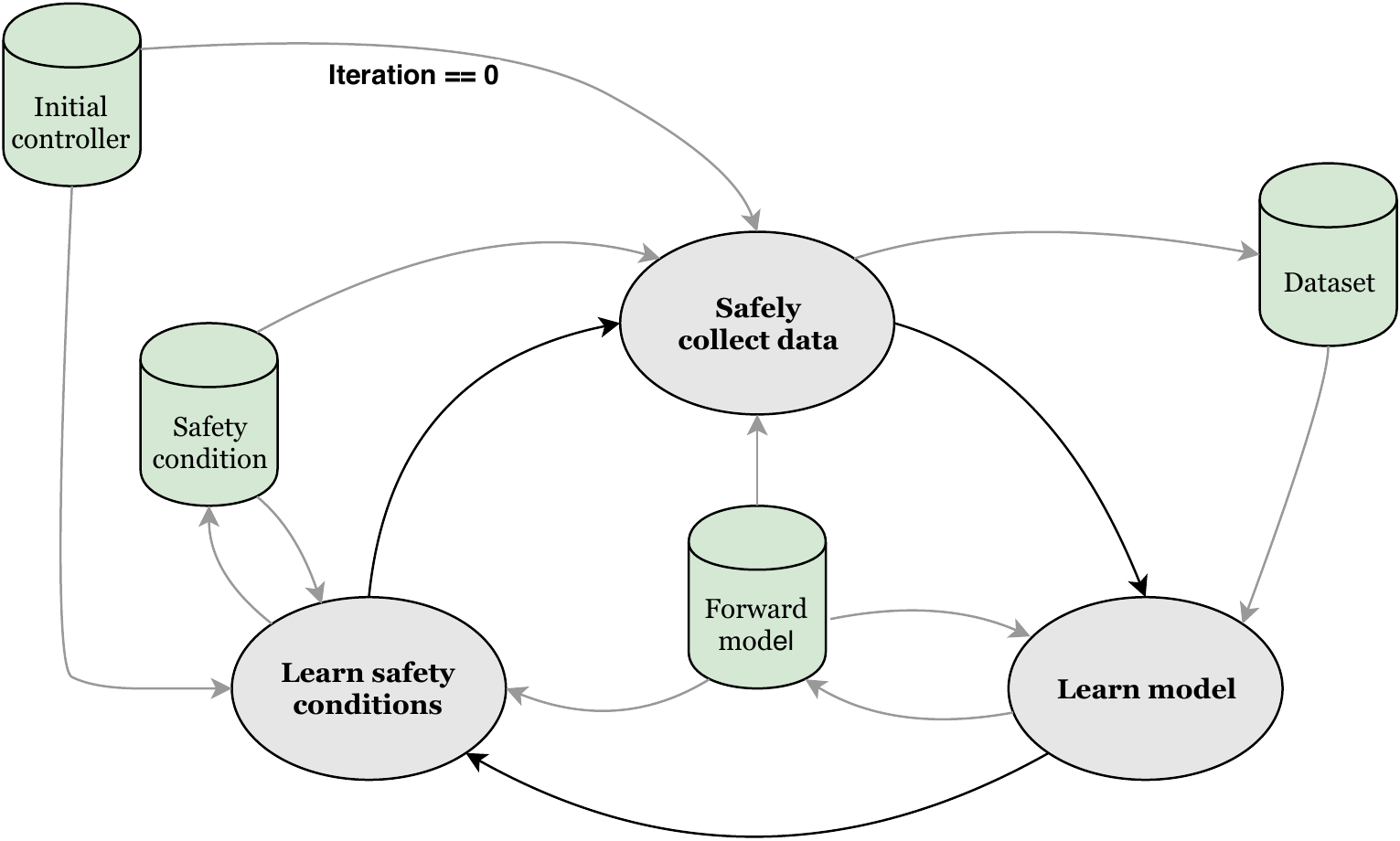}
       \caption{{\bf Safe interactive Model-Based Learning (SiMBL) rationale.} Approach is centered around an uncertain RNN forward model for which we compute a safe set and a control policy using principles from robust control. This allows for safe exploration though MPC and iterative refinement of the model and the safe set. An initial safe policy is assumed known. }    \label{fig:simbldiagram}   
      \vspace{-0.6cm}
\end{wrapfigure} The proposed framework aims at learning a policy, $K(x)$, and Lyapunov function, $V(x)$, 
by means of simulated trajectories from an uncertain forward model and an initial policy $K_0$, 
used to collect data. 
The model, the policy, and the Lyapunov function are \emph{iteratively refined} while \emph{safely} 
collecting more data 
 through a \emph{Safe Model Predictive Controller} (Safe-MPC).  
Figure~\ref{fig:simbldiagram} illustrates the approach. 

\paragraph{Summary of contribution.} This work presents the algorithms for: 
1) Iteratively learning a novel Bayesian RNN model with a large posterior over unseen states and inputs; 
2) Learning a safe set and the associated controller with 
neural networks from the model trajectories; 
3) Safe exploration with MPC. 
For 1) and 2), we propose to retain the model from scratch using a consistency prior to include knowledge 
of the previous uncertainty and then to recompute the safe set. 
The safe set increase as more data becomes available and the safety of the exploration strategy are 
demonstrated on an inverted pendulum simulation with limited control 
torque and stability region. Their final integration for continuous model and controller refinement with data from safe exploration (see Figure \ref{fig:simbldiagram}) is left for future work.

  \section{The Bayesian recurrent forward model} 
  A discrete-time stochastic forward model of system \eqref{eq:ode} is formulated as a Bayesian RNN. A grey-box approach is used, where available prior knowledge is integrated into the network in a differentiable way (for instance, the known relation between an observation and its derivative). The model provides an estimate of the next states distribution that is large (up to a defined value) where there is no available data. This is inspired by recent work on Noise Contrastive Priors (NCP) \citep{hafner_reliable_2018}.  We extend the NCP approach to RNNs, and propose the Noise Contrastive Prior Bayesian RNN (NCP-BRNN), with full state information, which follows the discrete-time update: 
\begin{align} \label{eq:forward}
      \hat{x}(t+1) = \hat{x}(t) + dt\ d\hat{x}(t),& \quad 
      d\hat{x}(t) = \mu\left(\hat{x}(t), u(t); \theta_\mu\right) + \hat{w}(t), \\
      \hat{w}(t)  \sim q(\hat{x}(t), u(t);\theta_\Sigma), &\quad q(\hat{x}(t), u(t);\theta_\Sigma) = \mathcal{N}\left(0,\ \Sigma\left(\hat{x}(t), u(t); \theta_\Sigma\right)\right), \\
      \hat{y}(t) \sim \mathcal{N}\left(\hat{x}(t), \sigma_y^2\right), \label{eq:measure} & \quad 
      \hat{x}(0) \sim \mathcal{N}(x(0), \sigma_y^2), \\
            \Sigma(\cdot) = \sigma_w \text{sigm}(\Sigma_{net}(\cdot)), &  \quad  \mu\left(\cdot\right)=\text{GreyBox}_{net}(\cdot), 
  \end{align}
  where $\hat{x}(t), \hat{y}(t)$ denote the state and measurement estimated from the model at time $t$,  and $d\hat{x}(t)$ is drawn from the distribution model, where $\mu$ and $\Sigma$ are computed from neural networks, sharing some initial layers. In particular, $\mu$ combines an MLP with some physics prior while the final activation of $\Sigma$ is a sigmoid which is then scaled by the hyperparameter $\sigma_w$, namely, a finite maximum variance. The next state distribution depends on the current state estimate $\hat{x}$, the input $u$, and a set of \emph{unknown} constant parameters $\theta$, which are to be learned from the data.  The estimated state $\hat{x}(t)$ is for simplicity assumed to have the same physical meaning of the true system state $x(t)$. The system state is measured with a Gaussian uncertainty with standard deviation $\sigma_y$, which is also learned from data. During the control, the measurement noise is assumed to be negligible ($\sigma_y\approx 0$). Therefore, the control algorithms will need to be robust with respect to the model uncertainty. Extensions to partial state information and output noise robust control are also possible but are left for future work.

\begin{wrapfigure}{l}{5.5cm}
\centering
         \vspace{0.cm}
       \includegraphics[width=1.05\linewidth]{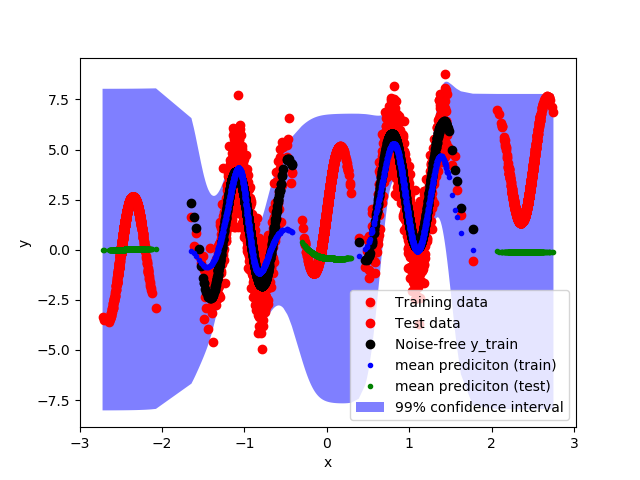}
       \caption{{\bf Variational neural networks with Noise Contrastive Priors (NCP).} Predicting sine-wave data (red-black) with confidence bounds (blue area) using NAIS-Net \citep{Ciccone2018NAISNetSD} and NCP \citep{hafner_reliable_2018}. }\label{fig:NCP}   
       \vspace{-1.3cm}
\end{wrapfigure}

\paragraph{Towards reliable uncertainty estimates with RNNs} 
The \emph{fundamental assumption} for model-based safe learning algorithm is that the model predictions \emph{contain} the actual system state transitions with high probability \citep{berkenkamp_safe_2017}. This is difficult to meet in practice for most neural network models. To mitigate this risk, we train our Bayesian RNNs on sequences and include a Noise-Contrastive Prior (NCP) term \citep{hafner_reliable_2018}. In the present work, the uncertainty is modelled as a point-wise Gaussian with mean and standard deviation that depend on the current state as well as on the input. The learned 1-step standard deviation, $\Sigma$, is assumed to be a diagonal matrix. This assumption is limiting but it is common in variational neural networks for practicality reasons \citep{zhao_infovae_2017, chen_variational_2016}. The NPC concept is illustrated in Figure \ref{fig:NCP}. More complex uncertainty representations will be considered in future works. 

The cost function used to train the model is:
\begin{align}\label{eq:model_loss}
    \mathcal{L}(\theta_\mu,\theta_\Sigma)  = & -\frac{1}{T}\sum_{t=0}^T
    \mathbf{E}_{p_{\text{train}}(x(0),u(t))}\left[
    \mathbf{E}_{q(\hat{x}(t), {u}(t);\theta_\Sigma)}\left[\ln p(\hat{y}(t)|y(t),u(t);\theta_\mu;\theta_\Sigma)\right]\right] \nonumber \\ & +  D_{\text{KL}}\left[q\left(\tilde{x}(t), \tilde{u}(t);\theta_\Sigma\right)\ ||\ \mathcal{N}(0,\sigma_w^2)\right] \nonumber \\ & +  \mathbf{E}_{p_{\text{train}}(x(0),u(t))}\left[\text{ReLU}\left[\Sigma\left(\hat{x}(t),u(t);\theta_{\Sigma}\right) - \Sigma\left(\hat{x}(t),u(t);\theta_{\Sigma_\text{prev}}\right)\right]\right]
\end{align}
where the first term is the expected negative log likelihood over the uncertainty distribution, evaluated over the training data. The second term is the KL-divergence which is evaluated in closed-form over predictions $\tilde{x}$ generated from a set of background initial states and input sequences, $\hat{x}(0)$ and $\hat{u}(t)$. These are sampled from a uniform distribution for the first model and then, once a previous model is available and new data is collected, they are obtained using rejection sampling with PyMC \citep{Salvatier2016} with acceptance condition: $\Sigma(\tilde{x}(0),\tilde{u}(t);\theta_{\Sigma_\text{prev}})\geq 0.5\  \sigma_w$. If a previous model is available, then the final term is used which is an \emph{uncertainty consistency prior} which forces the uncertainty estimates  over the training data to not increase with respect to the previous model.  
The loss (\ref{eq:model_loss}) is optimised using stochastic backpropagation through truncated sequences. In order to have further consistency between model updates, it a previous model is available, we train from scratch but stop optimising once the final loss of the previous model is reached. 

\section{The robust control problem} We approximate a chance-constrained stochastic control problem with a min-max robust control problem over a convex uncertainty set. This non-convex min-max control problem is then also approximated by computing the loss only at the vertices of the uncertainty set. To compensate for this approximation, (inspired by variational inference) the centre of the set is sampled from the uncertainty distribution itself (Figure \ref{fig:variational_sets}). 
The procedure is detailed below.

\paragraph{Lyapunov-Net.} The considered Lyapunov function is: 
\begin{equation}\label{eq:lyap1}
  V(x) = x^T\left(\epsilon I + V_{net}(x)^T V_{net}(x)\right)x + \psi(x),
  \end{equation}
  where $V_{net}(x)$ is a feedforward network that produces a  $n_V\times n_x$ matrix, where $n_V$ and $\epsilon>0$ are hyperparameters. The network parameters have to be trained and they are omitted from the notation. The term $\psi(x)\geq 0$ represents the prior knowledge of the state constraints. In this work we use:
  \begin{equation}
      \psi(x) = \text{ReLU}(\phi(x)-1),
  \end{equation}
  where $\phi(x)\geq 0$ is the Minkowski functional\footnote{Minkowski functionals measure the distance from the set center and they are positive definite.} of a user-defined \emph{usual region of operation}, namely:
  \begin{equation}
  \mathbb{X}_\phi=\{x \in \mathbb{X}: \phi(x)\leq 1\}.
  \end{equation}
  Possible choices for the Minkowski functional include quadratic functions, norms or semi-norms \citep{Blanchini,Horn:2012:MA:2422911}. 
  Since $V(x)$ must be positive definite, the hyperparameter $\epsilon>0$ is introduced~\footnote{The trainable part of the function $V(x)$ is chosen to be piece-wise quadratic but this is not the only possible choice. In fact one can use any positive definite and radially unbounded function. For the same problem multiple Lyapunov functions can exist. See also \cite{Blanchini}.}. While other forms are possible as in \cite{Blanchini}, with \eqref{eq:lyap1} the activation function does not need to be invertible. The study of the generality of the proposed function is left for future consideration. 

\paragraph{Safe set definition.} 
Denote the candidate \emph{safe level set} of $V$ as:
\begin{equation}
    \mathbb{X}_s = \{x \subseteq \mathbb{X}: V(x)\leq l_s\},
\end{equation}
where $l_s$ is the safe level. If, for $x \in \mathbb{X}_s$, the function $V(x)$ satisfies the Lyapunov inequality over the system closed-loop trajectory with a control policy $K$, namely, 
\begin{equation}\label{eq:lyap2}
    u(t)=K\left(x\left(t\right)\right)\Rightarrow V(x(t+1))-V(x(t))\leq 0,\ \forall x(t)\in \mathbb{X}_s,  
\end{equation} 
then set $\mathbb{X}_s$ is \emph{safe}, i.e., it satisfies the conditions of \emph{positive-invariance}  \citep{Blanchini,Kerrigan:2000}.  Note that the policy $K(x)$ can be either a neural network or a model-based controller, for instance a Linear Quadratic Regulator (LQR, see \cite{KalmanLQR}) or a Model Predictive Controller (MPC, see \cite{Maciejowski_book,rawlingsMPC,Cannon_book,Rakovic2019}). A stronger condition to eq. (\ref{eq:lyap2}) is often used in the context of optimal control: 
\begin{equation}\label{eq:lyap2_lqr}
    u(t)=K\left(x\left(t\right)\right)\Rightarrow V(x(t+1))-V(x(t))\leq -\ell(x(t),u(t)),\ \forall x(t)\in \mathbb{X}_s
\end{equation} 
where $\ell(x(t),u(t))$ is a positive semi-definite stage loss. 
In this paper, we focus on training policies with the quadratic loss used in LQR and MPC, where the origin is the target equilibrium, namely:
\begin{equation}
    \ell(x,u)= x^T Q x + u^T R u, \quad Q\succeq 0,\ R\succ 0. 
\end{equation}

\paragraph{From chance constrained to min-max control} 
Consider the problem of finding a controller $K$ and a function $V$ such that $u(t)=K\left(x\left(t\right)\right)$ and:
\begin{equation}
\label{eq:stochastic}
    \mathcal{P}\big[V(\hat{x}(t+1))-V(x(t))\leq -\ell(x(t),u(u))\big]\geq 1-\epsilon_p, 
\end{equation}
where $\mathcal{P}$ represents a probability and $0<\epsilon_p<<1$. This is a \emph{chance-constrained} non-convex optimal control problem \citep{Cannon_book}. We truncate the distributions and approximate (\ref{eq:stochastic}):
\begin{equation}
    \max_{\hat{x}(t+1) \in\  \mathbb{W}(x(t),u(t),\theta)}\big[V\left(\hat{x}(t+1)\right)\big]-V(x(t)) \leq - \ell\left(x(t), K\left(x(t)\right)\right), \label{eq:minmax_def}
\end{equation}
which is deterministic. A strategy to jointly learn $(V,\ K)$ fulfilling      (\ref{eq:minmax_def}) is presented next. 

\section{Learning the controller and the safe set}\label{sec:safeset}

We wish to build a controller $K$, a function $V$, and a safe level $l_s$ given the state transition probability model, $(\mu,\Sigma)$, such that the condition in (\ref{eq:lyap2_lqr}) is satisfied with high probability for the physical system generating the data. 
Denote the one-step prediction from the model in (\ref{eq:forward}), in closed loop with $K$, as:
$$\hat{x}^{+} = x + dt\ d\hat{x},\text{ with, }u=K(x),$$ 
 where $\hat{x}^{+}$ represents the next state prediction and the time index $t$ is omitted. 
 
 \paragraph{Approximating the high-confidence prediction set.} A polytopic approximation of a \emph{high confidence region} of the estimated uncertain set $\hat{x}^{+} \in  \mathbb{W}(x,u,\theta)$ is obtained from the parameters of $\Sigma$ and used for training $(V,\ K)$. In this work, the uncertain set is taken as a hyper-diamond centered at $x$, scaled by the (diagonal)  standard deviation matrix, $\Sigma$:
 \begin{equation}
    \mathbb{W}_1(x,u,\theta)=\left\{x^{+}: x^{+}=x + dt\ d\hat{x}, \ \left\| (\Sigma(x,u;\theta_\Sigma))^{-1}\hat{w}\right\|_1\leq \bar{\sigma} \right\},
 \end{equation}
 where $\bar{\sigma}>0$ is a hyper-parameter. This choice of set is inspired by the Unscented Kalman filter \citep{wan2000}.
 Since $\Sigma$ is diagonal, the vertices are given by the columns of the matrix resulting from multiplying $\Sigma$ with a mask $M$ such that:
\begin{align}
    \text{vert}[\mathbb{W}(x,u,\theta)]= \text{cols}[\Sigma(x,u;\theta_\Sigma)\ M], \quad 
    M= \bar{\sigma}\ [I,\ -I]. 
\end{align}

\paragraph{Learning the safe set.} Assume that a controller $K$ is given. Then, we wish to learn  a $V$ of the from of (\ref{eq:lyap1}), such that the corresponding safe set $\mathbb{X}_s$ is as big as possible, ideally as big as the state constraints $\mathbb{X}$. In order to do so, the parameters of $V_{net}$ are trained using a grid of initial states, a forward model to simulate the next state under the policy $K$, and an appropriate cost function.  The cost for $V_{net}$ and $l_s$ is inspired by \citep{richards_lyapunov_2018}. It consists of a combination of two objectives: the first one penalises the deviation from the Lyapunov stability condition; the second one is a classification penalty that separates the stable points from the unstable ones by means of the decision boundary, $V(x)=l_s$. The combined robust Lyapunov function cost is:
\begin{equation}\label{eq:lyapunov_loss}
    \min_{V_{net},\ l_s} \mathbf{E}_{[x(t)\in\mathbb{X}_{grid}]} \big[J\left(x\left(t\right)\right)\big], 
\end{equation}
\begin{align}\label{eq:cost}
J(x) &= \mathcal{I}_{\mathbb{X}_s}(x)\ J_{s}(x)+ \text{sign}\big[\nabla V(x)\big]\ 
\left[l_s - V(x) \right], 
\end{align}
\begin{align}
\mathcal{I}_{\mathbb{X}_s}(x) &= 0.5 \left(\text{sign}\left[l_s - V(x) \right]+1\right), \quad
J_{s}(x) = \frac{1}{\rho V(x)}\ \text{ReLU}\left[ \nabla V(x) \right], \\
\nabla V(x) &= \max_{\hat{x}^{+} \in\  \mathbb{W}(x,K\left(x\right),\theta)}\big[V\left(\hat{x}^{+}\right)\big]-V(x) + \ell\left(x, K\left(x\right)\right), \label{eq:minmax}
\end{align} 
where $\rho>0$ trades off stability for volume. The robust Lyapunov decrease in (\ref{eq:minmax}) is evaluated by using sampling to account for uncertainty over the confidence interval $\mathbb{W}$. Sampling of the set centre is performed as opposite of setting $\mathbb{W}=\mathbb{W}_1$, which didn't seem to produce valid results. Let us omit $\theta$ for ease of notation. We substitute $\nabla V(x)$ with $\mathbf{E}_{\mathbb{W}}\big[\nabla V(x)\big]$, which we define as: 
\begin{align}
\mathbf{E}_{\hat{w}  \sim \mathcal{N}\left(0,\ \Sigma\left(x, K(x)\right)\right)}\left\{\max_{\hat{x}^{+} \in\  \mathbb{W}_1(x,K\left(x\right),\theta) +  \hat{w}dt}\big[V\left(\hat{x}^{+}\right)\big]\right\}-V(x) + \ell\left(x, K\left(x\right)\right), \label{eq:expected:minmax}
 \end{align}

\begin{wrapfigure}{l}{5.5cm}
\vspace{-0.4cm}
    \centering
    \includegraphics[trim={120, 330, 150, 270}, clip, scale=0.55]{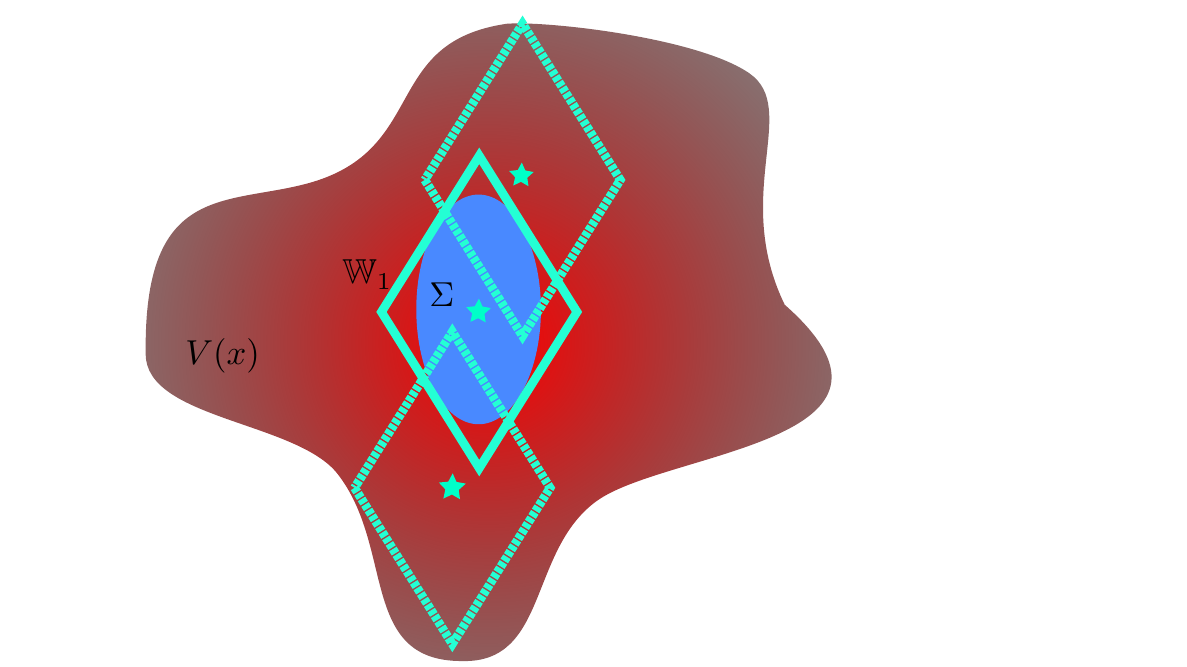}
    \caption{{\bf Approximating the non-convex maximisation.} Centre of the uncertain set is sampled and Lyapunov function is evaluated at its vertices. }
    \label{fig:variational_sets}
    \vspace{-0.3cm}
\end{wrapfigure}

Equations (\ref{eq:minmax}) and (\ref{eq:expected:minmax}) require a maximisation of the non-convex function $V(x)$ over the convex set $\mathbb{W}$. For the considered case, a sampling technique or another optimisation (similar to adversarial learning) could be used for a better approximation of the max operator. The maximum over $\mathbb{W}$ is instead approximated by the maximum over its vertices: 
\begin{equation}
\nabla V(x) \approx \max_{\hat{x}^{+} \in\ \text{vert}[\mathbb{W}_1(x,K\left(x\right),\theta)] +  \hat{w}dt}\big[V\left(\hat{x}^{+}\right)\big]-V(x) + \ell\left(x, K\left(x\right)\right). \label{eq:minmax_approx}
\end{equation}
This consists of a simple enumeration followed by a max over tensors that can be easily handled. Finally, during training (\ref{eq:expected:minmax}) is implemented in a variational inference  fashion by evaluating (\ref{eq:minmax_approx}) at each epoch over a different sample of $\hat{w}$. This entails a variational posterior over the center of the uncertainty interval. The approach is depicted in Figure \ref{fig:variational_sets}.  

The proposed cost is inspired by \cite{richards_lyapunov_2018}, with the difference that here there is no need for \emph{labelling} the states as safe by means of a multi-step simulation.  Moreover, in this work we train the Lyapunov function and controller \emph{together}, while in  \citep{richards_lyapunov_2018} the latter was given. 

\paragraph{Learning the safe policy.}  
 We alternate the minimisation of the Lyapunov loss (\ref{eq:lyapunov_loss}) and the solution of the \emph{variational robust control problem}:
\begin{align}\label{eq:policy_cost}
    \min_{u=K(x)}\mathbf{E}_{[x\in\mathbb{X}_{grid}]} \left[\mathcal{I}_{\mathbb{X}_s}(x) L_c(x,u)\right], \quad \text{s.t.: } K(0)=0,
\end{align}    
\begin{align}    
    L_c(x,u) = \ell(x,u) + &\mathbf{E}_{\mathbb{W}}\left\{\max_{\hat{x}^{+} \in\ \mathbb{W}(x,u,\theta)} \left[V(\hat{x}^{+}) - \gamma \log(l_s-V(\hat{x}^{+}))\right]\right\},  \label{eq:robust_control_loss}
\end{align}
subject to the forward model (\ref{eq:forward}). In this work,  (\ref{eq:policy_cost}) is solved using backpropagation through the policy, the model and $V$. The safety constraint, $\hat{x}^{+}\in\mathbb{X}_s$, namely, $V(\hat{x}^{+})\leq l_s$ is relaxed through a log-barrier \citep{Boyd:2004:CO:993483}. If a neural policy $K(x)$ solves (\ref{eq:policy_cost}) and satisfies the safety constraint, $\forall x\in\mathbb{X}_s$, then it is a candidate robust controller for keeping the system within the safe set $\mathbb{X}_s$. Note that the expectation in (\ref{eq:robust_control_loss}) is once again treated as a variational approximation of the expectation over the center of the uncertainty interval.  

Obtaining an exact solution to the control problem for all points is computationally impractical. In order to provide statistical guarantees of safety, probabilistic verification is used after $V$ and $K$ have been trained.  This refines the safe level set $l_s$ and, if successful, provides a probabilistic safety certificate. If the verification is unsuccessful, then the learned $(\mathbb{X}_s,\ K)$ are not safe. The data collection continues with the previous safe controller until suitable $V$, $l_s$, and $K$ are found. Note that the number of training points used for the safe set and controller is in general lower than the ones used for verification. The alternate learning procedure for $\mathbb{X}_s$ and $K$ is summarised in Algorithm \ref{alg:alternateDescent}. The use of 1-step predictions makes the procedure highly scalable through parallelisation on GPU. 


\begin{wrapfigure}{r}{6.5cm}
    \noindent\begin{minipage}{0.45\columnwidth}
    \vspace{-0.1cm}
        \begin{algorithm}[H]
          \DontPrintSemicolon
            \KwInput{$K_0$, $\mathbb{X}_{\text{grid}}$, $\theta_\mu$, $\theta_\Sigma$, $\sigma_w\geq0$, $\bar{\sigma}>0$, $\epsilon>0$}
            \KwOutput{$(V_{net},\ l_s,\ K)$}
            \small
            \caption{Alternate descent for safe set}
            \label{alg:alternateDescent}
            \For{$i=0...N$}{
                \For{$j=0...N_v$}{
                    $(V_{net},\ l_s)\leftarrow$ Adam step on (\ref{eq:cost}) \;
                }
                \For{$j=0...N_k$}{
                    $K \leftarrow$ Adam step on (\ref{eq:policy_cost}) \;
                }
            }
        \end{algorithm}
    \end{minipage}
    \vspace{-0.4cm}
 \end{wrapfigure}

\paragraph{Probabilistic safety verification.}
A probabilistic verification is used to numerically prove the physical system stability with high probability. The resulting certificate is of the form (\ref{eq:stochastic}), where the $\epsilon_P$ decreases with increasing number of samples. Following the work of  \cite{bobiti_samplingdriven_nodate}, the simulation is evaluated at a large set of points within the estimated safe set $\mathbb{X}_s$. Monte Carlo rejection sampling is performed with PyMC \citep{Salvatier2016}. 
\begin{wrapfigure}{r}{7.5cm}
\vspace{-0.4cm}
    \noindent\begin{minipage}{0.53\columnwidth}
        \begin{algorithm}[H]
            \DontPrintSemicolon
            \KwInput{$N$, $V$, $K$, $\theta_\mu$, $\theta_\Sigma$, $\sigma_w\geq0$, $\bar{\sigma}>0$, $\delta>0$}
            \KwOutput{$(\text{SAFE},\ l_u,\  l_l)$}
            \small
            \caption{Probabilistic safety verification}
            \label{alg:verification}
            $\text{SAFE}\leftarrow \text{False}$ \;
            \For{$l_u = 1,1-\delta,1-2\delta,...,0$ }{
                \For{$l_l = 0, \delta, 2\delta,...,l_u$}{
                    draw $N$ uniform $x$-samples s.t.:\\\qquad $l_l\ l_s\leq V(x)\leq l_u\ l_s$ \;
                    draw $N$ $w$-samples from $\mathcal{U}(-\bar{\sigma},\bar{\sigma})$ \;
                    \If{$V(\hat{x}^{+})-V(x)\leq 0,\forall x,\forall w$}{
                        draw $N$ uniform $x$-samples s.t.:\\\qquad $V(x)\leq l_l\ l_s$ \;
                        \If{$V(\hat{x}^{+})\leq l_u\ l_s,\forall x,\forall w$}{
                            $\text{SAFE}\leftarrow \text{True}$ \;
                            \textbf{return} \text{SAFE}, $l_u$, $l_l$\;
                        }
                    }
                }
            }
            Verification failed.\;  
        \end{algorithm}
    \end{minipage}
    \vspace{-0.5cm}       
\end{wrapfigure}

In practical applications, several factors limit the convergence of the trajectory to a neighborhood of the target (the \emph{ultimate bound}, \cite{Blanchini}). For instance,  the policy structural bias, discount factors in RL methods or persistent uncertainty in the model, the state estimates, and the physical system itself. Therefore, we extended the verification algorithm of \citep{bobiti_samplingdriven_nodate} to estimate the ultimate bound as well as the invariant set, as outlined in Algorithm \ref{alg:verification}. Given a maximum and minimum level, $l_l$, $l_u$, we first sample initial states uniformly within these two levels and check for a robust decrease of $V$ over the next state distribution. If this is verified, then we sample uniformly from inside the minimum level set $l_l$ (where $V$ may not decrease) and check that $V$ does not exceed the maximum level $l_u$ over the next state distribution. The distribution is evaluated by means of uniform samples of $w$, independent of the current state, within $(-\bar{\sigma},\bar{\sigma})$. These are then scaled using $\Sigma$ from the model. We search for $l_l$, $l_u$ with a step $\delta$.

Note that, in Algorithm \ref{alg:verification}, the uncertainty of the surrogate model is taken into account by sampling a single uncertainty realisation for the entire set of initial states. The values of $w$ will be then scaled using $\Sigma$ in the forward model. This step is computationally convenient but breaks the assumption that variables are drawn from a uniform distribution. We leave this to future work. In this paper, independent Gaussian uncertainty models are used and stability is verified directly on the environment. Note that probabilistic verification is expensive but necessary, as pathological cases could result in the training loss (\ref{eq:lyapunov_loss}) for the safe set could converging to a local minima with a very small set. If this is the case then usually the forward model is not accurate enough or the uncertainty hyperparameter $\sigma_w$ is too large. Note that  Algorithm \ref{alg:verification} is highly parallelizable. 

\section{Safe exploration}\label{sec:exploration}
Once a verified safe set is found the environment can be controlled by means of a 1-step MPC with probabilistic stability (see Appendix). Consider the constraint $V(x)\leq l^{\star}_s=l_u\ l_s$, where $V$ and $l_s$ come from Algorithm \ref{alg:alternateDescent} and $l_u$ from Algorithm \ref{alg:verification}. The Safe-MPC exploration strategy follows: 
\paragraph{Safe-MPC for exploration.} For collecting new data, solve the following MPC problem: 
\begin{eqnarray}\label{eq:MPCexploration1}
u^\star= \arg\min_{u\in\mathbb{U}}\bigg\{ \beta\ell(x,u) -\alpha\ \ell_{expl}(x,u) + \max_{\hat{x}^{+} \in\ \mathbb{W}(x,u,\theta)}\left[\beta V(\hat{x}^{+}) - \gamma \log(l^{\star}_s-V(\hat{x}^{+}))\right]\bigg\}, 
\end{eqnarray}
where $\alpha\leq\gamma$ is the \emph{exploration} hyperparameter, $\beta\in[0,1]$ is the \emph{regulation} or \emph{exploitation} parameter and $\ell_{expl}(x,u)$ is the info-gain from the model, similar to \citep{hafner_reliable_2018}:
\begin{equation}
    \ell_{expl}(x,u) = \sum_{i=1,\dots, N_x}\frac{\left(\Sigma_{ii}\left({x}, u, p, \theta\right)\right)^2}{N_x\ \sigma_{y_{ii}}^2}.  
\end{equation}
The full derivation of the problem and a probabilistic safety result are discussed in Appendix. 
\paragraph{Alternate min-max optimization.}
Problem (\ref{eq:MPCexploration1}) is approximated using alternate descent. In particular, the maximization in the loss function over the uncertain future state $\hat{x}^{+}$ with respect to $\hat{w}$, given the current control candidate $u$, is alternated with the minimization with respect to $u$, given the current candidate $\hat{x}^{+}$. Adam \citep{kingma_adam:_2014} is used for both steps.

\section{Inverted pendulum example}
The approach is demonstrated on an inverted pendulum, where the input is the angular torque and the states/outputs are the angular position and velocity of the pendulum. The aim is to collect data safely around the unstable equilibrium point (the origin). The system has a torque constraint that limits the controllable region. In particular, if the initial angle is greater than 60 degrees, then the torque is not sufficient to swing up the pendulum. In order to compare to the LQR, we choose a linear policy with a $\tanh$ activation, meeting the torque constraints while preserving differentiability. 

\begin{figure}[h!]
    \centering
    \begin{subfigure}[b]{0.195\textwidth}
    \centering
        \includegraphics[trim={45 23.5 0 0}, clip, scale=0.245]{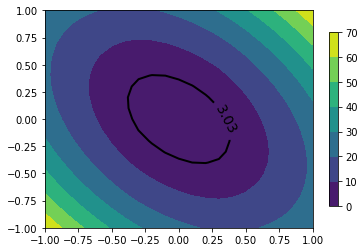}
            \caption*{\centering{\hspace*{1.5em} \tiny $i=1$\newline $K=[-10,  -0.05]$}}
    \end{subfigure}
    \begin{subfigure}[b]{0.195\textwidth}
    \centering
        \includegraphics[trim={45 23.5 0 0}, clip, scale=0.245]{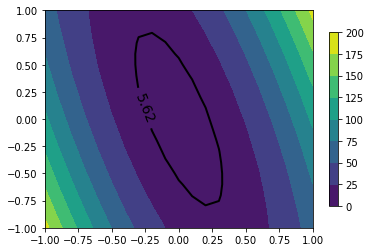}
            \caption*{\centering{\hspace*{1.5em} \tiny $i=30$\newline $K=[-9.24,\ -1.56]$}}
    \end{subfigure}
    \begin{subfigure}[b]{0.195\textwidth}
    \centering
        \includegraphics[trim={45 23.5 0 0}, clip, scale=0.245]{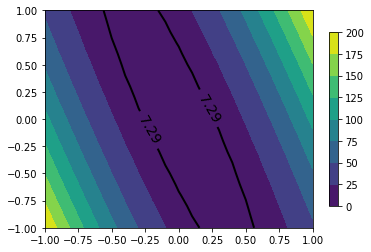}
            \caption*{\centering{\hspace*{1.5em} \tiny $i=50$\newline $K=[-8.49,\ -2.25]$}}
    \end{subfigure}
        \begin{subfigure}[b]{0.195\textwidth}
    \centering
        \includegraphics[trim={45 23.5 0 0}, clip, scale=0.245]{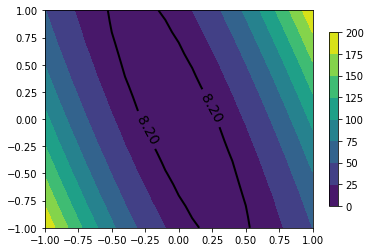}
            \caption*{\centering{\hspace*{1.5em} \tiny $i=60$\newline $K=[-8.1, -2.7]$}}
    \end{subfigure}
    \begin{subfigure}[b]{0.195\textwidth}
    \centering
        \includegraphics[trim={45 23.5 0 20}, clip, scale=0.245]{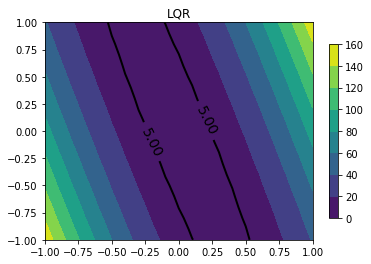}
        \caption*{\centering{\hspace*{1.5em} \tiny LQR\newline $K=[-7.26,\ -2.55]$}}
    \end{subfigure}
    \caption{{\bf Inverted Pendulum. Safe set and controller with proposed method for known environment model.} Initial set ($i=0$) is based on a unit circle plus the constraint $|\alpha|\leq 0.3$. Contours show the function levels. Control gain gets closer to the LQR solution as iterations progress until circa $i=50$, where the minimum of the Lyapunov loss (\ref{eq:lyapunov_loss}) is achieved. The set and controller at iteration $50$ are closest to the LQR solution, which is optimal around the equilibrium in the unconstrained case. In order to maximise the chances of verification the optimal parameters are selected with a early stopping, namely when the Lyapunov loss reaches its minimum, resulting in $K=[-8.52,\ -2.2]$. }
    \label{fig:pendulum_lyap_nom}
\end{figure}

\paragraph{Safe set with known environment, comparison to LQR.}
We first test the safe-net algorithm on the nominal pendulum model and compare the policy and the safe set with those from a standard LQR policy. Figure \ref{fig:pendulum_lyap_nom} shows the safe set at different stages of the algorithm, approaching the LQR.  

\begin{figure}[h!]
    \centering
    \begin{subfigure}[b]{0.19\textwidth}
        \includegraphics[trim={52 35 101 45}, clip, scale=0.224]{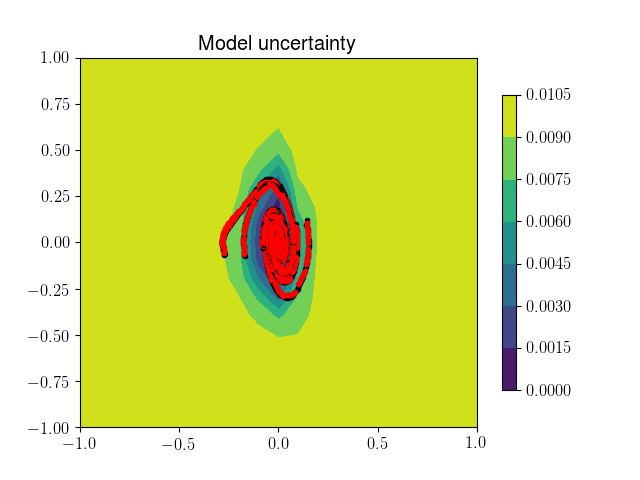}\vspace{-0.1cm}
        \caption*{$10k$ points}
    \end{subfigure}
    \begin{subfigure}[b]{0.19\textwidth}
    \centering
        \includegraphics[trim={52 35 101 45}, clip, scale=0.224]{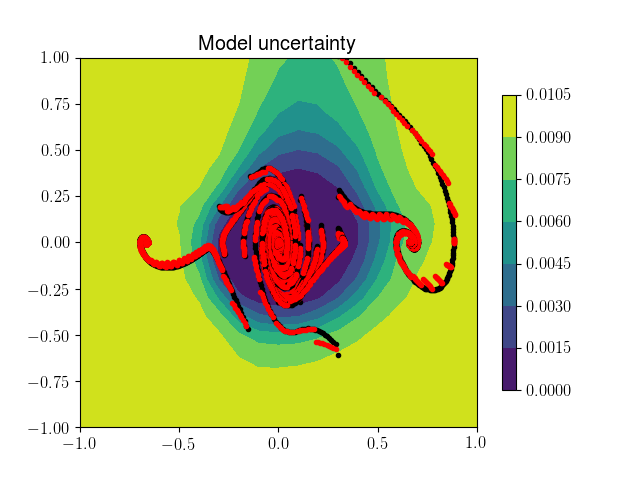}\vspace{-0.1cm}
            \caption*{$30k$ points}
    \end{subfigure}
    \begin{subfigure}[b]{0.19\textwidth}
    \centering
        \includegraphics[trim={52 35 101 45}, clip, scale=0.224]{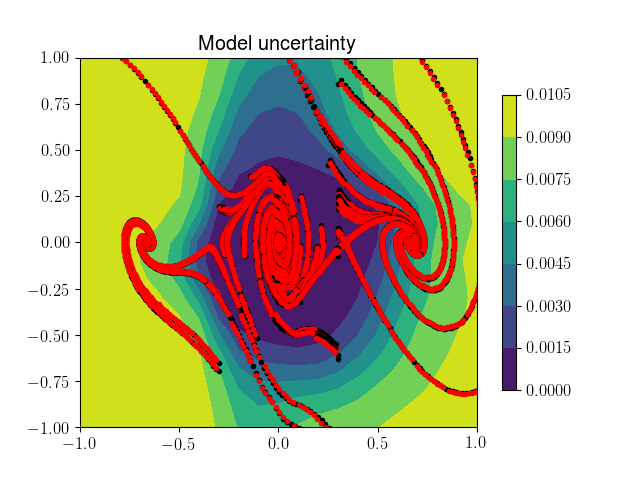}\vspace{-0.1cm}
            \caption*{$50k$ points}
    \end{subfigure}
    \begin{subfigure}[b]{0.19\textwidth}
    \centering
        \includegraphics[trim={52 35 101  45}, clip, scale=0.224]{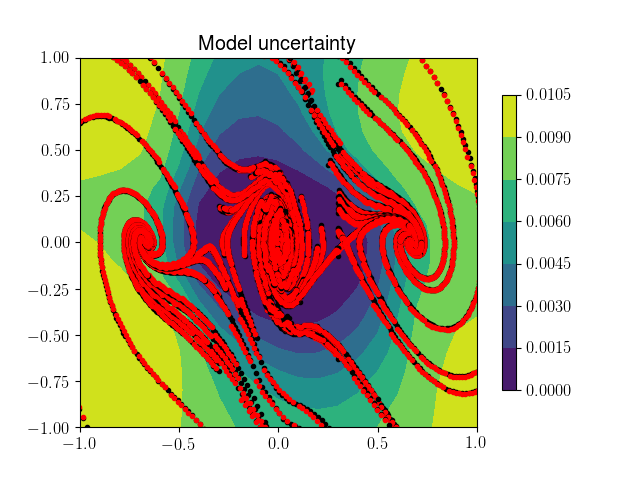}\vspace{-0.1cm}
            \caption*{$70k$ points}
    \end{subfigure}
        \begin{subfigure}[b]{0.19\textwidth}
    \centering
        \includegraphics[trim={52 35 50 45}, clip, scale=0.224]{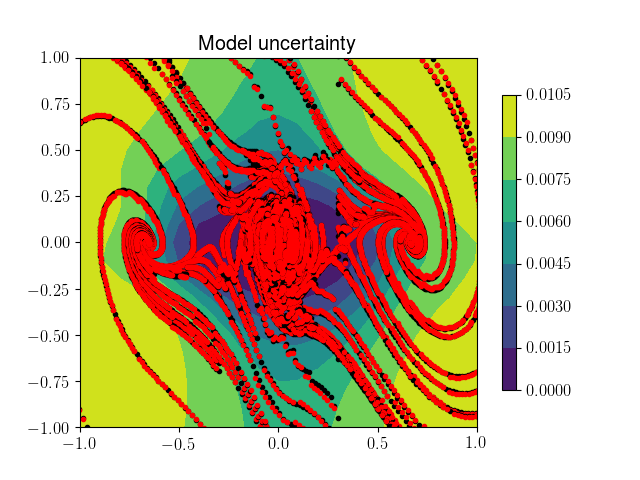}\vspace{-0.1cm}
            \caption*{$90k$ points}
    \end{subfigure}
 
    \begin{subfigure}[b]{0.1965\textwidth}
    \centering
        \includegraphics[trim={52 35 65 45}, clip, scale=0.21]{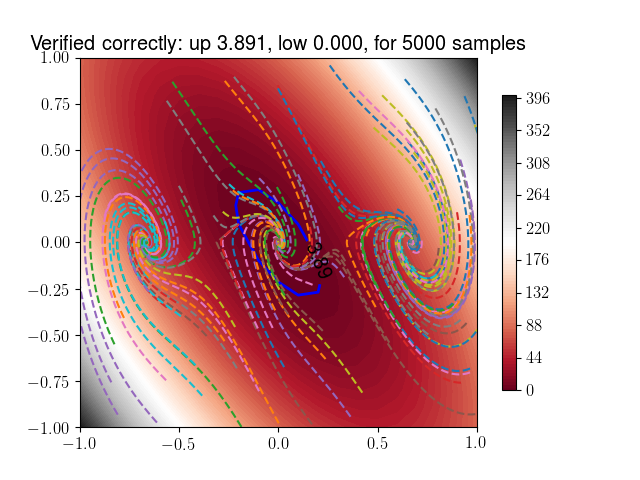}\vspace{-0.1cm}
        \caption*{$10k$ points}
    \end{subfigure}
    \begin{subfigure}[b]{0.195\textwidth}
    \centering
        \includegraphics[trim={52 35 65 45}, clip, scale=0.21]{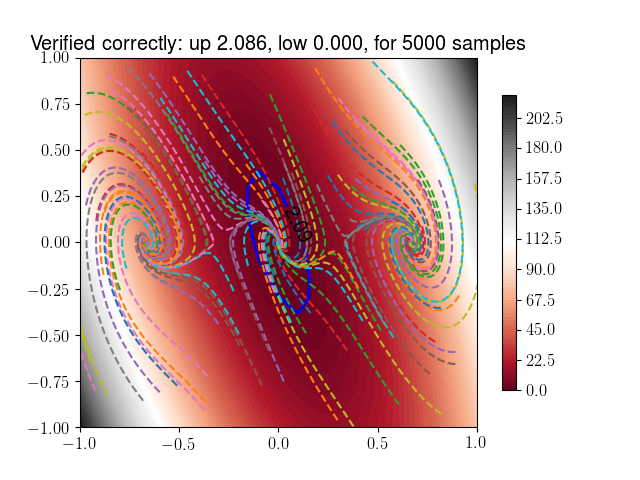}\vspace{-0.1cm}
            \caption*{$30k$ points}
    \end{subfigure}
    \begin{subfigure}[b]{0.195\textwidth}
    \centering
        \includegraphics[trim={52 35 65 45}, clip, scale=0.21]{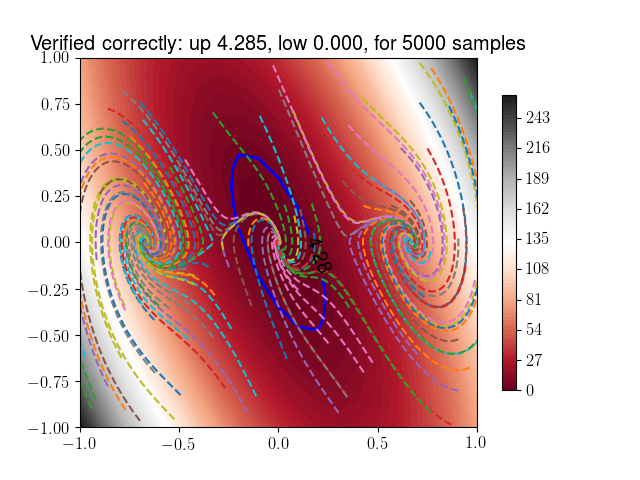}\vspace{-0.1cm}
            \caption*{$50k$ points}
    \end{subfigure}
    \begin{subfigure}[b]{0.195\textwidth}
    \centering
        \includegraphics[trim={52 35 65 45}, clip, scale=0.21]{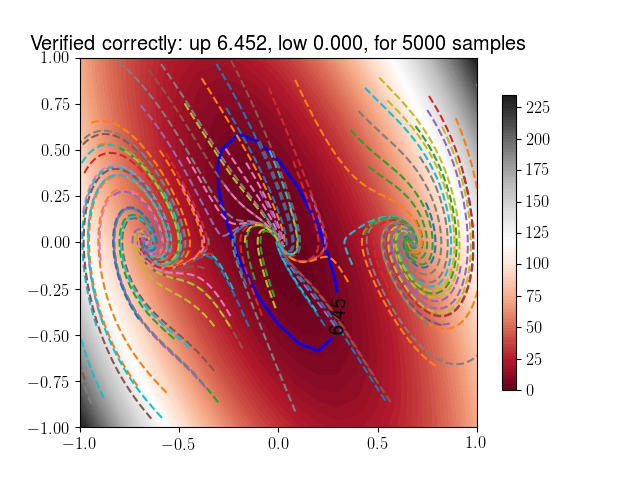}\vspace{-0.1cm}
            \caption*{$80k$ points}
    \end{subfigure}
        \begin{subfigure}[b]{0.195\textwidth}
    \centering
        \includegraphics[trim={52 35 65 45}, clip, scale=0.21]{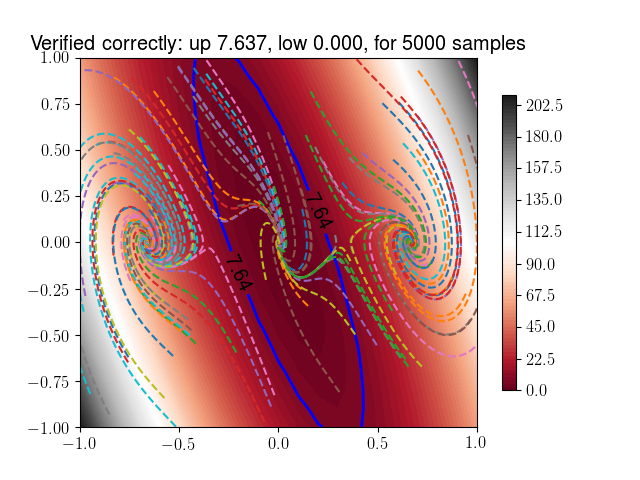}\vspace{-0.1cm}
            \caption*{environment}
    \end{subfigure}
    \caption{{\bf Inverted pendulum safe set with Bayesian model.} Surrogates are obtained with increasing amount of data. The initial state and input perturbation from the safe policy are drawn from Gaussians with standard deviation that doubles each $10k$ points. {\bf Top}: Mean predictions and uncertainty contours for the NCP-BRNN model.  After $90k$ points no further improvement is noticed. {\bf Bottom}: Comparison of safe sets with surrogates and environment. Reducing the model uncertainty while maintaining a similar prediction accuracy leads to an increase in the safe set. After $90k$ points no further benefits are noticed on the set which is consistent with the uncertainty estimates.     }
    \label{fig:pendulum_lyap}
\end{figure}

\vspace{-0.5cm} 
\paragraph{Safe set with Bayesian model.} In order to test the proposed algorithms, the forward model is fitted on sequences of length $10$ for an increasing amount of data points ($10k$ to $100k$). Data is collected in closed loop with the initial controller, $K_0=[-10,\ 0]$, with different initial states. In particular, we perturb the initial state and control values with a random noise with standard deviations starting from, respectively, $0.1$ and $0.01$ and doubling each $10k$ points. The only prior used is that the velocity is the derivative of the angular position (normalized to $2\pi$ and $\pi$). The uncertainty bound was fixed to $\sigma_w=0.01$. The architecture was cross-validated from $60k$ datapoints with a $70$-$30$ split. The model with the best validation predictions as well as the largest safe set was used to generate the results in Figure \ref{fig:pendulum_lyap}. 
\begin{wrapfigure}{l}{7.5cm}
\vspace{-0.3cm}
    \centering
    \begin{subfigure}[b]{0.22\textwidth}
    \centering
        \includegraphics[trim={52 35 0 0}, clip, scale=0.25]{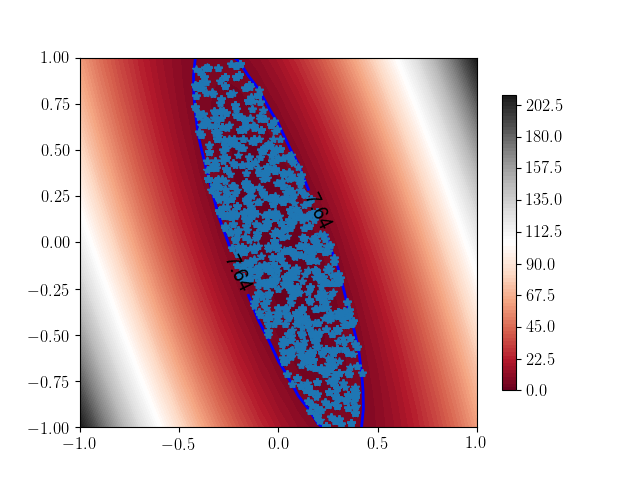}
        \caption{\scriptsize environment}
    \end{subfigure}
    \begin{subfigure}[b]{0.22\textwidth}
    \centering
        \includegraphics[trim={52 35 0 0}, clip, scale=0.25]{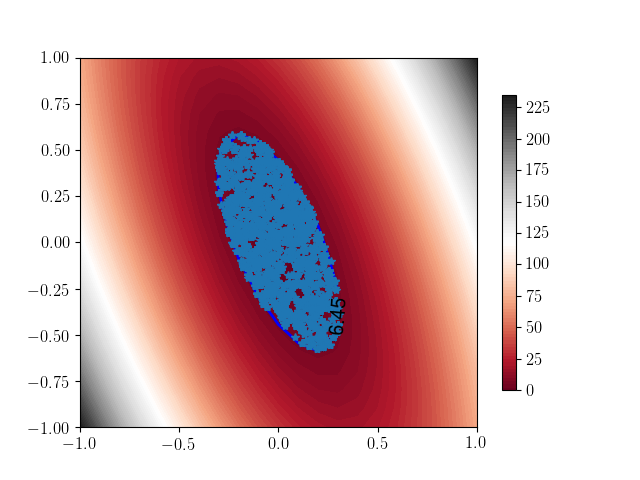}
            \caption{\scriptsize NCP-BRNN ($80k$ points).}
    \end{subfigure}
    \caption{{\bf Inverted pendulum verification}. Nominal and robust safe sets are verified on the pendulum simulation using $5k$ samples. We search for the largest stability region and the smallest ultimate bound of the solution. If a simulation is not available, then a two-level sampling on BRNN is performed.}
    \label{fig:verification}
  \vspace{-0.9cm} 
\end{wrapfigure}
The results demonstrate that the size of the safe set can improve with more data, provided that the model uncertainty decreases and the predictions have comparable accuracy. This motivates for exploration.

\vspace{-0.1cm}

\paragraph{Verification on the environment.} 
The candidate Lyapunov function, safe level set, and robust control policy are formally verified through probabilistic sampling of the system state, according to Algorithm \ref{alg:verification}, where the simulation is used directly. The results for $5k$ samples are shown in Figure \ref{fig:verification}. In particular, the computed level sets verify at the first attempt and no further search for sub-levels or ultimate bounds is needed. 

\begin{wrapfigure}{r}{7.5cm}
    \centering
    \begin{subfigure}[b]{0.25\textwidth}
    \centering
        \includegraphics[trim={40 23.5 0 20}, clip, scale=0.3]{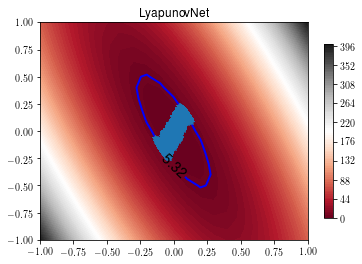}
        \caption*{\centering{\hspace*{1em} \scriptsize Semi-random exploration\newline \hspace*{2em} $50$ trials of $1k$ steps\newline  $\text{vol}=0.06$}}
    \end{subfigure}
    \begin{subfigure}[b]{0.25\textwidth}
    \centering
        \includegraphics[trim={44.5 26 0 23.5}, clip, scale=0.3]{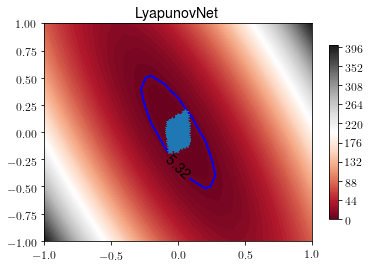}
            \caption*{\centering{\hspace*{1em} \scriptsize Safe-MPC exploration\newline \hspace*{2em} $1$ trial of $50k$ steps\newline  $\text{vol}=0.04$}}
    \end{subfigure}
    \caption{{\bf Safe exploration}. Comparison of a naive semi-random exploration strategy with the proposed Safe-MPC for exploration. The proposed algorithm has an efficient space coverage with safety guarantees. }
    \label{fig:pendulum_explore}
\end{wrapfigure} 

\paragraph{Safe exploration.}

Safe exploration is performed using the min-max approach in Section \ref{sec:exploration}. For comparison, a semi-random exploration strategy is also used: if inside the safe set, the action magnitude is set to maximum torque and its sign is given by a random uniform variable  once $V(x)\geq 0.99\ l_s$, then the safe policy $K$ is used. This does not provide any formal guarantees of safety as the value of $V(x)$ could exceed the safe level, especially for very fast systems and large input signals. This is repeated for several trials in order to estimate the maximum reachable set within the safe set. The results are shown in Figure \ref{fig:pendulum_explore}, where the semi-random strategy is used as a baseline and is compared to a single trial of the proposed safe-exploration algorithm. The area covered by our algorithm in a \emph{single trial} of $50k$ steps is about $67\%$ of that of the semi-random baseline over $50$ trials of $1k$ steps. Extending the length of the trials did not significantly improve the baseline results. Despite being more conservative, our algorithm continues to explore safely indefinitely.




\section{Conclusions}
Preliminary results show that the SiMBL produces a Lyapunov function and a safe set using neural networks that are comparable with that of standard optimal control (LQR) and can account for state-dependant additive model uncertainty. A Bayesian RNN surrogate with NCP was proposed and trained for an inverted pendulum simulation. An alternate descent method was presented to jointly learn a Lyapunov function, a safe level set, and a stabilising control policy for the surrogate model with back-propagation. We demonstrated that adding data-points to the training set can increase the safe-set size provided that the model improves and its uncertainty decreases. To this end, an uncertainty prior from the previous model was added to the framework. The safe set was then formally verified through a novel probabilistic algorithm for ultimate bounds and used for safe data collection (exploration). A one-step safe MPC was proposed where the Lyapunov function provides the terminal cost and constraint to mimic an infinite horizon with high probability of recursive feasibility. Results show that the proposed safe-exploration strategy has better coverage than a naive policy which switches between random inputs and the safe policy. 

\bibliographystyle{apa-good}
\bibliography{./simbl.bib}

\begin{thebibliography}{65}
\expandafter\ifx\csname natexlab\endcsname\relax\def\natexlab#1{#1}\fi
\expandafter\ifx\csname url\endcsname\relax
  \def\url#1{{\tt #1}}\fi
\expandafter\ifx\csname urlprefix\endcsname\relax\def\urlprefix{URL }\fi

\bibitem[{Akametalu et~al.(2014)Akametalu, Fisac, Gillula, Kaynama, Zeilinger,
  \& Tomlin}]{Akametalu2014}
Akametalu, A.~K., Fisac, J.~F., Gillula, J.~H., Kaynama, S., Zeilinger, M.~N.,
  \& Tomlin, C.~J. (2014).
\newblock Reachability-based safe learning with gaussian processes.
\newblock In {\em 53rd {IEEE} Conference on Decision and Control\/}. {IEEE}.
\newline\urlprefix\url{https://doi.org/10.1109/cdc.2014.7039601}

\bibitem[{Bemporad et~al.(2003)Bemporad, Borrelli, \& Morari}]{Bemporad_minmax}
Bemporad, A., Borrelli, F., \& Morari, M. (2003).
\newblock Min-max control of constrained uncertain discrete-time linear
  systems.
\newblock {\em Automatic Control, IEEE Transactions on\/}, {\em 48\/}, 1600 --
  1606.

\bibitem[{Ben-Tal et~al.(2009)Ben-Tal, Ghaoui, \& Nemirovski}]{Ben-Tal}
Ben-Tal, A., Ghaoui, L.~E., \& Nemirovski, A. (2009).
\newblock {\em Robust Optimization (Princeton Series in Applied
  Mathematics)\/}.
\newblock Princeton University Press.

\bibitem[{Berkenkamp et~al.(2017)Berkenkamp, Turchetta, Schoellig, \&
  Krause}]{berkenkamp_safe_2017}
Berkenkamp, F., Turchetta, M., Schoellig, A.~P., \& Krause, A. (2017).
\newblock Safe {Model}-based {Reinforcement} {Learning} with {Stability}
  {Guarantees}.
\newblock {\em arXiv:1705.08551 [cs, stat]\/}.
\newblock ArXiv: 1705.08551.
\newline\urlprefix\url{http://arxiv.org/abs/1705.08551}

\bibitem[{Blanchini \& Miani(2007)}]{Blanchini}
Blanchini, F., \& Miani, S. (2007).
\newblock {\em Set-Theoretic Methods in Control (Systems \& Control:
  Foundations \& Applications)\/}.
\newblock Birkhäuser.

\bibitem[{Bobiti \& Lazar(2016)}]{bobiti_sampling-based_2016}
Bobiti, R., \& Lazar, M. (2016).
\newblock Sampling-based verification of {Lyapunov}'s inequality for piecewise
  continuous nonlinear systems.
\newblock {\em arXiv:1609.00302 [cs]\/}.
\newblock ArXiv: 1609.00302.
\newline\urlprefix\url{http://arxiv.org/abs/1609.00302}

\bibitem[{Bobiti(2017)}]{bobiti_samplingdriven_nodate}
Bobiti, R.~V. (2017).
\newblock {\em Sampling driven stability domains computation and predictive
  control of constrained nonlinear systems\/}.
\newblock Ph.D. thesis.
\newline\urlprefix\url{https://pure.tue.nl/ws/files/78458403/20171025_Bobiti.pdf}

\bibitem[{Borrelli et~al.(2017)Borrelli, Bemporad, \& Morari}]{Borrelli_book}
Borrelli, F., Bemporad, A., \& Morari, M. (2017).
\newblock {\em Predictive Control for Linear and Hybrid Systems\/}.
\newblock Cambridge University Press.

\bibitem[{Boyd \& Vandenberghe(2004)}]{Boyd:2004:CO:993483}
Boyd, S., \& Vandenberghe, L. (2004).
\newblock {\em Convex Optimization\/}.
\newblock New York, NY, USA: Cambridge University Press.

\bibitem[{Camacho \& Bordons(2007)}]{Camacho2007}
Camacho, E.~F., \& Bordons, C. (2007).
\newblock {\em Model Predictive control\/}.
\newblock Springer London.

\bibitem[{Carron et~al.(2019)Carron, Arcari, Wermelinger, Hewing, Hutter, \&
  Zeilinger}]{Carron2019}
Carron, A., Arcari, E., Wermelinger, M., Hewing, L., Hutter, M., \& Zeilinger,
  M.~N. (2019).
\newblock Data-driven model predictive control for trajectory tracking with a
  robotic arm.
\newline\urlprefix\url{http://hdl.handle.net/20.500.11850/363021}

\bibitem[{Chen et~al.(2016)Chen, Kingma, Salimans, Duan, Dhariwal, Schulman,
  Sutskever, \& Abbeel}]{chen_variational_2016}
Chen, X., Kingma, D.~P., Salimans, T., Duan, Y., Dhariwal, P., Schulman, J.,
  Sutskever, I., \& Abbeel, P. (2016).
\newblock Variational {Lossy} {Autoencoder}.
\newblock {\em arXiv:1611.02731 [cs, stat]\/}.
\newblock ArXiv: 1611.02731.
\newline\urlprefix\url{http://arxiv.org/abs/1611.02731}

\bibitem[{Cheng et~al.(2019)Cheng, Orosz, Murray, \&
  Burdick}]{cheng_end--end_2019}
Cheng, R., Orosz, G., Murray, R.~M., \& Burdick, J.~W. (2019).
\newblock End-to-{End} {Safe} {Reinforcement} {Learning} through {Barrier}
  {Functions} for {Safety}-{Critical} {Continuous} {Control} {Tasks}.
\newblock {\em arXiv:1903.08792 [cs, stat]\/}.
\newblock ArXiv: 1903.08792.
\newline\urlprefix\url{http://arxiv.org/abs/1903.08792}

\bibitem[{Chow et~al.(2018)Chow, Nachum, Duenez-Guzman, \&
  Ghavamzadeh}]{chow_lyapunov-based_2018}
Chow, Y., Nachum, O., Duenez-Guzman, E., \& Ghavamzadeh, M. (2018).
\newblock A {Lyapunov}-based {Approach} to {Safe} {Reinforcement} {Learning}.
\newblock {\em arXiv:1805.07708 [cs, stat]\/}.
\newblock ArXiv: 1805.07708.
\newline\urlprefix\url{http://arxiv.org/abs/1805.07708}

\bibitem[{Chow et~al.(2019)Chow, Nachum, Faust, Duenez-Guzman, \&
  Ghavamzadeh}]{chow_lyapunov-based_2019}
Chow, Y., Nachum, O., Faust, A., Duenez-Guzman, E., \& Ghavamzadeh, M. (2019).
\newblock Lyapunov-based {Safe} {Policy} {Optimization} for {Continuous}
  {Control}.
\newblock {\em arXiv:1901.10031 [cs, stat]\/}.
\newblock ArXiv: 1901.10031.
\newline\urlprefix\url{http://arxiv.org/abs/1901.10031}

\bibitem[{Chua et~al.(2018)Chua, Calandra, McAllister, \&
  Levine}]{chua_deep_2018}
Chua, K., Calandra, R., McAllister, R., \& Levine, S. (2018).
\newblock Deep {Reinforcement} {Learning} in a {Handful} of {Trials} using
  {Probabilistic} {Dynamics} {Models}.
\newblock {\em arXiv:1805.12114 [cs, stat]\/}.
\newblock ArXiv: 1805.12114.
\newline\urlprefix\url{http://arxiv.org/abs/1805.12114}

\bibitem[{Ciccone et~al.(2018)Ciccone, Gallieri, Masci, Osendorfer, \&
  Gomez}]{Ciccone2018NAISNetSD}
Ciccone, M., Gallieri, M., Masci, J., Osendorfer, C., \& Gomez, F. (2018).
\newblock Nais-net: Stable deep networks from non-autonomous differential
  equations.
\newblock In {\em NeurIPS\/}.

\bibitem[{Deisenroth \& Rasmussen(2011)}]{DeisenrothRT2011}
Deisenroth, M., \& Rasmussen, C. (2011).
\newblock Pilco: A model-based and data-efficient approach to policy search.
\newblock In {\em Proceedings of the 28th International Conference on Machine
  Learning, ICML 2011\/}, (pp. 465--472). Omnipress.

\bibitem[{Deisenroth et~al.(2015)Deisenroth, Fox, \&
  Rasmussen}]{deisenroth_gaussian_2015}
Deisenroth, M.~P., Fox, D., \& Rasmussen, C.~E. (2015).
\newblock Gaussian {Processes} for {Data}-{Efficient} {Learning} in {Robotics}
  and {Control}.
\newblock {\em IEEE Transactions on Pattern Analysis and Machine
  Intelligence\/}, {\em 37\/}(2), 408--423.
\newblock ArXiv: 1502.02860.
\newline\urlprefix\url{http://arxiv.org/abs/1502.02860}

\bibitem[{Depeweg et~al.(2016)Depeweg, Hernández-Lobato, Doshi-Velez, \&
  Udluft}]{depeweg_learning_2016}
Depeweg, S., Hernández-Lobato, J.~M., Doshi-Velez, F., \& Udluft, S. (2016).
\newblock Learning and {Policy} {Search} in {Stochastic} {Dynamical} {Systems}
  with {Bayesian} {Neural} {Networks}.
\newblock {\em arXiv:1605.07127 [cs, stat]\/}.
\newblock ArXiv: 1605.07127.
\newline\urlprefix\url{http://arxiv.org/abs/1605.07127}

\bibitem[{Frigola et~al.(2014)Frigola, Chen, \&
  Rasmussen}]{Frigola2014VariationalGP}
Frigola, R., Chen, Y., \& Rasmussen, C.~E. (2014).
\newblock Variational gaussian process state-space models.
\newblock In {\em NIPS\/}.

\bibitem[{Gal et~al.(2016)Gal, McAllister, \& Rasmussen}]{Gal2016Improving}
Gal, Y., McAllister, R., \& Rasmussen, C.~E. (2016).
\newblock Improving {PILCO} with {B}ayesian neural network dynamics models.
\newblock In {\em Data-Efficient Machine Learning workshop, ICML\/}.

\bibitem[{Gallieri(2016)}]{Gallieri2016}
Gallieri, M. (2016).
\newblock {\em Lasso-{MPC} {\textendash} Predictive Control with
  $\ell_1$-Regularised Least Squares\/}.
\newblock Springer International Publishing.
\newline\urlprefix\url{https://doi.org/10.1007/978-3-319-27963-3}

\bibitem[{Gros \& Zanon(2019)}]{gros_towards_2019}
Gros, S., \& Zanon, M. (2019).
\newblock Towards {Safe} {Reinforcement} {Learning} {Using} {NMPC} and {Policy}
  {Gradients}: {Part} {II} - {Deterministic} {Case}.
\newblock {\em arXiv:1906.04034 [cs]\/}.
\newblock ArXiv: 1906.04034.
\newline\urlprefix\url{http://arxiv.org/abs/1906.04034}

\bibitem[{Hafner et~al.(2018{\natexlab{a}})Hafner, Lillicrap, Fischer,
  Villegas, Ha, Lee, \& Davidson}]{hafner_learning_2018}
Hafner, D., Lillicrap, T., Fischer, I., Villegas, R., Ha, D., Lee, H., \&
  Davidson, J. (2018{\natexlab{a}}).
\newblock Learning {Latent} {Dynamics} for {Planning} from {Pixels}.
\newblock {\em arXiv:1811.04551 [cs, stat]\/}.
\newblock ArXiv: 1811.04551.
\newline\urlprefix\url{http://arxiv.org/abs/1811.04551}

\bibitem[{Hafner et~al.(2018{\natexlab{b}})Hafner, Tran, Irpan, Lillicrap, \&
  Davidson}]{hafner_reliable_2018}
Hafner, D., Tran, D., Irpan, A., Lillicrap, T., \& Davidson, J.
  (2018{\natexlab{b}}).
\newblock Reliable {Uncertainty} {Estimates} in {Deep} {Neural} {Networks}
  using {Noise} {Contrastive} {Priors}.
\newblock {\em arXiv:1807.09289 [cs, stat]\/}.
\newblock ArXiv: 1807.09289.
\newline\urlprefix\url{http://arxiv.org/abs/1807.09289}

\bibitem[{Hewing et~al.(2017)Hewing, Kabzan, \&
  Zeilinger}]{hewing_cautious_2017}
Hewing, L., Kabzan, J., \& Zeilinger, M.~N. (2017).
\newblock Cautious {Model} {Predictive} {Control} using {Gaussian} {Process}
  {Regression}.
\newblock {\em arXiv:1705.10702 [cs, math]\/}.
\newblock ArXiv: 1705.10702.
\newline\urlprefix\url{http://arxiv.org/abs/1705.10702}

\bibitem[{Horn \& Johnson(2012)}]{Horn:2012:MA:2422911}
Horn, R.~A., \& Johnson, C.~R. (2012).
\newblock {\em Matrix Analysis\/}.
\newblock New York, NY, USA: Cambridge University Press, 2nd ed.

\bibitem[{Kalman(2001)}]{KalmanLQR}
Kalman, R. (2001).
\newblock Contribution to the theory of optimal control.
\newblock {\em Bol. Soc. Mat. Mexicana\/}, {\em 5\/}.

\bibitem[{Kerrigan(2000)}]{Kerrigan:2000}
Kerrigan, E. (2000).
\newblock Robust constraint satisfaction: Invariant sets and predictive
  control.
\newblock Tech. rep.
\newline\urlprefix\url{http://hdl.handle.net/10044/1/4346}

\bibitem[{Kerrigan \& Maciejowski(2004)}]{Kerrigan2004}
Kerrigan, E.~C., \& Maciejowski, J.~M. (2004).
\newblock Feedback min-max model predictive control using a single linear
  program: robust stability and the explicit solution.
\newblock {\em International Journal of Robust and Nonlinear Control\/}, {\em
  14\/}(4), 395--413.
\newline\urlprefix\url{https://doi.org/10.1002/rnc.889}

\bibitem[{Khalil(2014)}]{Khalil_book}
Khalil, H.~K. (2014).
\newblock {\em Nonlinear Control\/}.
\newblock Pearson.

\bibitem[{Kingma \& Ba(2014)}]{kingma_adam:_2014}
Kingma, D.~P., \& Ba, J. (2014).
\newblock Adam: {A} {Method} for {Stochastic} {Optimization}.
\newblock {\em arXiv:1412.6980 [cs]\/}.
\newblock ArXiv: 1412.6980.
\newline\urlprefix\url{http://arxiv.org/abs/1412.6980}

\bibitem[{Koller et~al.(2018)Koller, Berkenkamp, Turchetta, \&
  Krause}]{koller_learning-based_2018}
Koller, T., Berkenkamp, F., Turchetta, M., \& Krause, A. (2018).
\newblock Learning-based {Model} {Predictive} {Control} for {Safe}
  {Exploration} and {Reinforcement} {Learning}.
\newblock {\em arXiv:1803.08287 [cs]\/}.
\newblock ArXiv: 1803.08287.
\newline\urlprefix\url{http://arxiv.org/abs/1803.08287}

\bibitem[{Kouvaritakis \& Cannon(2015)}]{Cannon_book}
Kouvaritakis, B., \& Cannon, M. (2015).
\newblock {\em {Model Predictive Control: Classical, Robust and Stochastic}\/}.
\newblock Advanced Textbooks in Control and Signal Processing, Springer,
  London.

\bibitem[{Kurutach et~al.(2018)Kurutach, Clavera, Duan, Tamar, \&
  Abbeel}]{kurutach_model-ensemble_2018}
Kurutach, T., Clavera, I., Duan, Y., Tamar, A., \& Abbeel, P. (2018).
\newblock Model-{Ensemble} {Trust}-{Region} {Policy} {Optimization}.
\newblock {\em arXiv:1802.10592 [cs]\/}.
\newblock ArXiv: 1802.10592.
\newline\urlprefix\url{http://arxiv.org/abs/1802.10592}

\bibitem[{Limon et~al.(2017)Limon, Calliess, \& Maciejowski}]{Limon2017}
Limon, D., Calliess, J., \& Maciejowski, J. (2017).
\newblock Learning-based nonlinear model predictive control.
\newblock {\em {IFAC}-{PapersOnLine}\/}, {\em 50\/}(1), 7769--7776.
\newline\urlprefix\url{https://doi.org/10.1016/j.ifacol.2017.08.1050}

\bibitem[{Lorenzen et~al.(2019)Lorenzen, Cannon, \& Allgower}]{lorenzen_cannon}
Lorenzen, M., Cannon, M., \& Allgower, F. (2019).
\newblock {Robust MPC with recursive model update}.
\newblock {\em Automatica\/}, {\em 103\/}, 467--471.
\newline\urlprefix\url{https://ora.ox.ac.uk/objects/pubs:965898}

\bibitem[{Lowrey et~al.(2018)Lowrey, Rajeswaran, Kakade, Todorov, \&
  Mordatch}]{lowrey_plan_2018}
Lowrey, K., Rajeswaran, A., Kakade, S., Todorov, E., \& Mordatch, I. (2018).
\newblock Plan {Online}, {Learn} {Offline}: {Efficient} {Learning} and
  {Exploration} via {Model}-{Based} {Control}.
\newblock {\em arXiv:1811.01848 [cs, stat]\/}.
\newblock ArXiv: 1811.01848.
\newline\urlprefix\url{http://arxiv.org/abs/1811.01848}

\bibitem[{Maciejowski(2000)}]{Maciejowski_book}
Maciejowski, J. (2000).
\newblock {\em Predictive Control with Constraints\/}.
\newblock Prentice Hall.

\bibitem[{Mayne et~al.(2000)Mayne, Rawlings, Rao, \&
  Scokaert}]{rawlings_mayne_paper}
Mayne, D.~Q., Rawlings, J.~B., Rao, C.~V., \& Scokaert, P. O.~M. (2000).
\newblock Constrained model predictive control: {Stability} and optimality.

\bibitem[{Papini et~al.(2018)Papini, Battistello, Restelli, \&
  Battistello}]{Papini2018SafelyEP}
Papini, M., Battistello, A., Restelli, M., \& Battistello, A. (2018).
\newblock Safely exploring policy gradient.

\bibitem[{Pozzoli(2019)}]{pozzoli}
Pozzoli, S. (2019).
\newblock {\em State Estimation and Recurrent Neural Networks for Model
  Predictive Control\/}.
\newblock Politecnico di Milano, MS thesis, supervisors: R. Scattolini, M.
  Gallieri, E. Terzi, M. Farina.

\bibitem[{Pozzoli et~al.(2019)Pozzoli, Gallieri, \&
  Scattolini}]{pozzoli_tustin_2019}
Pozzoli, S., Gallieri, M., \& Scattolini, R. (2019).
\newblock Tustin neural networks: a class of recurrent nets for adaptive {MPC}
  of mechanical systems.
\newblock {\em arXiv:1911.01310 [cs, eess]\/}.
\newblock ArXiv: 1911.01310.
\newline\urlprefix\url{http://arxiv.org/abs/1911.01310}

\bibitem[{Raimondo et~al.(2009)Raimondo, Limon, Lazar, Magni, \&
  Camacho}]{Raimondo2009}
Raimondo, D., Limon, D., Lazar, M., Magni, L., \& Camacho, E. (2009).
\newblock Min-max model predictive control of nonlinear systems: A unifying
  overview on stability.
\newblock {\em European Journal of Control\/}, {\em 15\/}.

\bibitem[{Rakovi{\'{c}} et~al.(2012)Rakovi{\'{c}}, Kouvaritakis, Findeisen, \&
  Cannon}]{Rakovic2012}
Rakovi{\'{c}}, S.~V., Kouvaritakis, B., Findeisen, R., \& Cannon, M. (2012).
\newblock Homothetic tube model predictive control.
\newblock {\em Automatica\/}, {\em 48\/}, 1631--1638.

\bibitem[{Rakovi{\'{c}} \& Levine(2019)}]{Rakovic2019}
Rakovi{\'{c}}, S.~V., \& Levine, W.~S. (Eds.)  (2019).
\newblock {\em Handbook of Model Predictive Control\/}.
\newblock Springer International Publishing.
\newline\urlprefix\url{https://doi.org/10.1007/978-3-319-77489-3}

\bibitem[{Rawlings \& Mayne(2009)}]{rawlingsMPC}
Rawlings, J.~B., \& Mayne, D.~Q. (2009).
\newblock {\em Model Predictive Control Theory and Design\/}.
\newblock Nob Hill Pub, Llc.

\bibitem[{Richards(2004)}]{richards_a._g._robust_2004}
Richards, A.~G. (2004).
\newblock {\em Robust {Constrained} {Model} {Predictive} {Control} ,\/}.
\newblock Ph.D. thesis, MIT.

\bibitem[{Richards et~al.(2018)Richards, Berkenkamp, \&
  Krause}]{richards_lyapunov_2018}
Richards, S.~M., Berkenkamp, F., \& Krause, A. (2018).
\newblock The {Lyapunov} {Neural} {Network}: {Adaptive} {Stability}
  {Certification} for {Safe} {Learning} of {Dynamical} {Systems}.
\newblock {\em arXiv:1808.00924 [cs]\/}.
\newblock ArXiv: 1808.00924.
\newline\urlprefix\url{http://arxiv.org/abs/1808.00924}

\bibitem[{Salimans et~al.(2017)Salimans, Ho, Chen, Sidor, \&
  Sutskever}]{salimans2017es}
Salimans, T., Ho, J., Chen, X., Sidor, S., \& Sutskever, I. (2017).
\newblock Evolution strategies as a scalable alternative to reinforcement
  learning.
\newblock {\em arXiv preprint arXiv:1703.03864\/}.

\bibitem[{Salvatier et~al.(2016)Salvatier, Wiecki, \&
  Fonnesbeck}]{Salvatier2016}
Salvatier, J., Wiecki, T.~V., \& Fonnesbeck, C. (2016).
\newblock Probabilistic programming in python using {PyMC}3.
\newblock {\em {PeerJ} Computer Science\/}, {\em 2\/}, e55.
\newline\urlprefix\url{https://doi.org/10.7717/peerj-cs.55}

\bibitem[{Shyam et~al.(2018)Shyam, Jaskowski, \& Gomez}]{max}
Shyam, P., Jaskowski, W., \& Gomez, F. (2018).
\newblock Model-based active exploration.
\newblock {\em CoRR\/}, {\em abs/1810.12162\/}.
\newline\urlprefix\url{http://arxiv.org/abs/1810.12162}

\bibitem[{Stanley \& Miikkulainen(2002)}]{stanley2002evolving}
Stanley, K.~O., \& Miikkulainen, R. (2002).
\newblock Evolving neural networks through augmenting topologies.
\newblock {\em Evolutionary computation\/}, {\em 10\/}(2), 99--127.

\bibitem[{Taylor et~al.(2019)Taylor, Dorobantu, Le, Yue, \&
  Ames}]{taylor_episodic_2019}
Taylor, A.~J., Dorobantu, V.~D., Le, H.~M., Yue, Y., \& Ames, A.~D. (2019).
\newblock Episodic {Learning} with {Control} {Lyapunov} {Functions} for
  {Uncertain} {Robotic} {Systems}.
\newblock {\em arXiv:1903.01577 [cs]\/}.
\newblock ArXiv: 1903.01577.
\newline\urlprefix\url{http://arxiv.org/abs/1903.01577}

\bibitem[{Thananjeyan et~al.(2019)Thananjeyan, Balakrishna, Rosolia, Li,
  McAllister, Gonzalez, Levine, Borrelli, \&
  Goldberg}]{thananjeyan_safety_2019}
Thananjeyan, B., Balakrishna, A., Rosolia, U., Li, F., McAllister, R.,
  Gonzalez, J.~E., Levine, S., Borrelli, F., \& Goldberg, K. (2019).
\newblock Safety {Augmented} {Value} {Estimation} from {Demonstrations}
  ({SAVED}): {Safe} {Deep} {Model}-{Based} {RL} for {Sparse} {Cost} {Robotic}
  {Tasks}.
\newblock {\em arXiv:1905.13402 [cs, stat]\/}.
\newblock ArXiv: 1905.13402.
\newline\urlprefix\url{http://arxiv.org/abs/1905.13402}

\bibitem[{Verdier \& M.~Mazo(2017)}]{verdier_formal_2017}
Verdier, C.~F., \& M.~Mazo, J. (2017).
\newblock Formal {Controller} {Synthesis} via {Genetic} {Programming}.
\newblock {\em IFAC-PapersOnLine\/}, {\em 50\/}(1), 7205--7210.
\newline\urlprefix\url{https://doi.org/10.1016/j.ifacol.2017.08.1362}

\bibitem[{Vinogradska(2017)}]{vinogradska_gaussian_nodate}
Vinogradska, J. (2017).
\newblock Gaussian {Processes} in {Reinforcement} {Learning}: {Stability}
  {Analysis} and {Efficient} {Value} {Propagation}.
\newline\urlprefix\url{http://tuprints.ulb.tu-darmstadt.de/7286/1/GPs_in_RL_Stability_Analysis_and_Efficient_Value_Propagation_Version1.pdf}

\bibitem[{Wabersich et~al.(2019)Wabersich, Hewing, Carron, \&
  Zeilinger}]{wabersich_probabilistic_2019}
Wabersich, K.~P., Hewing, L., Carron, A., \& Zeilinger, M.~N. (2019).
\newblock Probabilistic model predictive safety certification for
  learning-based control.
\newblock {\em arXiv:1906.10417 [cs, eess]\/}.
\newblock ArXiv: 1906.10417.
\newline\urlprefix\url{http://arxiv.org/abs/1906.10417}

\bibitem[{Wan \& {Van Der Merwe}(2000)}]{wan2000}
Wan, E., \& {Van Der Merwe}, R. (2000).
\newblock The unscented kalman filter for nonlinear estimation.
\newblock (pp. 153--158).

\bibitem[{Williams et~al.(2017)Williams, Wagener, Goldfain, Drews, Rehg, Boots,
  \& Theodorou}]{Williams2017}
Williams, G., Wagener, N., Goldfain, B., Drews, P., Rehg, J.~M., Boots, B., \&
  Theodorou, E.~A. (2017).
\newblock Information theoretic {MPC} for model-based reinforcement learning.
\newblock In {\em 2017 {IEEE} International Conference on Robotics and
  Automation ({ICRA})\/}. {IEEE}.
\newline\urlprefix\url{https://doi.org/10.1109/icra.2017.7989202}

\bibitem[{Yan et~al.(2018)Yan, Goulart, \& Cannon}]{yan_stochastic_2018}
Yan, S., Goulart, P., \& Cannon, M. (2018).
\newblock Stochastic {Model} {Predictive} {Control} with {Discounted}
  {Probabilistic} {Constraints}.
\newblock ArXiv: 1807.07465.
\newline\urlprefix\url{http://arxiv.org/abs/1807.07465}

\bibitem[{Yang \& Maciejowski(2015{\natexlab{a}})}]{Yang2015b}
Yang, X., \& Maciejowski, J. (2015{\natexlab{a}}).
\newblock Risk-sensitive model predictive control with gaussian process models.
\newblock {\em {IFAC}-{PapersOnLine}\/}, {\em 48\/}(28), 374--379.
\newline\urlprefix\url{https://doi.org/10.1016/j.ifacol.2015.12.156}

\bibitem[{Yang \& Maciejowski(2015{\natexlab{b}})}]{Yang2015}
Yang, X., \& Maciejowski, J.~M. (2015{\natexlab{b}}).
\newblock Fault tolerant control using gaussian processes and model predictive
  control.
\newblock {\em International Journal of Applied Mathematics and Computer
  Science\/}, {\em 25\/}(1), 133--148.
\newline\urlprefix\url{https://doi.org/10.1515/amcs-2015-0010}

\bibitem[{Zhao et~al.(2017)Zhao, Song, \& Ermon}]{zhao_infovae_2017}
Zhao, S., Song, J., \& Ermon, S. (2017).
\newblock {InfoVAE}: {Information} {Maximizing} {Variational} {Autoencoders}.
\newblock {\em arXiv:1706.02262 [cs, stat]\/}.
\newblock ArXiv: 1706.02262.
\newline\urlprefix\url{http://arxiv.org/abs/1706.02262}

\end{thebibliography}

\newpage
\appendix

\section{Robust optimal control for safe learning}
Further detail is provided regarding robust and chance constrained control. 

\paragraph{Chance-constrained and robust control.} 
Consider the problem of finding a controller $K$ and a function $V$ such that $u(t)=K\left(x\left(t\right)\right)$ and:
\begin{equation}
\label{eq:stochastic2}
    \mathcal{P}\big[V(\hat{x}(t+1))-V(x(t))\leq -\ell(x(t),u(u))\big]\geq 1-\epsilon_p, 
\end{equation}

where $\hat{x}$ is given by the forward model (\ref{eq:forward}), $\mathcal{P}$ represents a probability and $0<\epsilon_p<<1$. This is a \emph{chance-constrained} control problem \citep{Cannon_book,yan_stochastic_2018}. Since finding $K$ and $V$ that satisfy (\ref{eq:stochastic2}) requires solving a non-convex and also stochastic optimization, we approximate (\ref{eq:stochastic2}) with a min-max condition over a high-confidence interval, in the form of a convex set $\mathbb{W}(x(t),u(t),\theta)$, as follows:
\begin{equation}
    \max_{\hat{x}(t+1) \in\  \mathbb{W}(x(t),u(t),\theta)}\big[V\left(\hat{x}(t+1)\right)\big]-V(x(t)) \leq - \ell\left(x(t), K\left(x(t)\right)\right), \label{eq:minmax_def2}
\end{equation}
This is a robust control problem, which is still non-convex but deterministic. In the convex case, (\ref{eq:minmax_def2}) can be satisfied by means of robust optimization \citep{Ben-Tal,rawlingsMPC}. By following this consideration, we frame the control problem as a non-convex min-max optimization. 

\paragraph{Links to optimal control and intrinsic robustness.}
To link our approach with optimal control and reinforcement learning, note that if the condition in (\ref{eq:lyap2_lqr}) is met with equality, then the controller $K$ and the Lyapunov function $V$ satisfy the Bellman equation \citep{rawlingsMPC}. Therefore, $u=K(x)$ is optimal and $V(x)$ is the value-function of the infinite horizon optimal control problem with stage loss $\ell(x,u)$. In practice, this condition is not met with exact equality. Nevertheless, the inequality in (\ref{eq:lyap2_lqr}) guarantees by definition that the system controlled by $K(x)$ is asymptotically stable (converges to $x=0$) and it has a degree of tolerance to uncertainty in the safe set $\mathbb{X}_s$ (i.e. if the system is locally Lipschitz)  \citep{rawlingsMPC}. Vice versa, infinite horizon optimal control with the considered cost produces a value function which is also a Lyapunov function and provides an intrinsic degree of robustness \citep{rawlingsMPC}. 

\section{From robust MPC to safe exploration}\label{sec:exploration2}
Once a robust Lyapunov function and invariant set are found, the environment can be controlled by means of a one-step MPC with probabilistic safety guarantees. 
\paragraph{One-step robust MPC.}
Start by considering the following min-max 1-step MPC problem:
\begin{align}\label{eq:MPC_hard}
u^\star & = \arg\min_{u\in\mathbb{U}}\bigg\{ \ell(x,u) + \max_{\hat{x}^{+} \in\ \mathbb{W}(x,u,\theta)}\left[V(\hat{x}^{+})\right]\bigg\}, \\ 
& \text{s.t.} \max_{\hat{x}^{+} \in\ \mathbb{W}(x,u,\theta)}\left[ V(\hat{x}^{+})\right]\leq l_s,  \text{ and to (\ref{eq:forward}) -- (\ref{eq:measure})}, \nonumber\\
& \text{given} \quad x=x(t),\ \text{ with } V(x)\leq l_s. \nonumber
\end{align}
This is a non-convex min-max optimisation problem with hard non-convex constraints. Solving (\ref{eq:MPC_hard}) is difficult, especially in real-time, but is in general possible if the constraints are feasible. This is true with a probability that depends from the verification procedure, the confidence level used in the procedures, as well as the probability of the model being correct.  

\paragraph{Relaxed  problem.} Solutions of (\ref{eq:MPC_hard}) can be computed in real-time, to a degree of accuracy, by iterative convexification of the problem and the use of fast convex solvers. This is described in Appendix. For the purpose of this paper, we will consider the \emph{soft-constrained} or \emph{relaxed} problem: 
\begin{eqnarray}\label{eq:MPC}
u^\star= \arg\min_{u\in\mathbb{U}}\bigg\{ \ell(x,u) + \max_{\hat{x}^{+} \in\ \mathbb{W}(x,u,\theta)}\left[V(\hat{x}^{+}) - \gamma \log(l_s-V(\hat{x}^{+}))\right]\bigg\}, 
\end{eqnarray}
once again subject to (\ref{eq:forward}). It is assumed that a  scalar, $\gamma>0$, exists such that the constraint can be enforced. For the sake of simplicity, problem (\ref{eq:MPC}) will be addressed using backpropagation, at the price of losing real-time guarantees. 

\paragraph{Safe exploration.} For collecting new data, we modify the robust MPC problem as follows:
\begin{eqnarray}\label{eq:MPCexploration}
u^\star= \arg\min_{u\in\mathbb{U}}\bigg\{ \beta\ell(x,u) -\alpha\ \ell_{expl}(x,u) + \max_{\hat{x}^{+} \in\ \mathbb{W}(x,u,\theta)}\left[\beta V(\hat{x}^{+}) - \gamma \log(l_s-V(\hat{x}^{+}))\right]\bigg\}, 
\end{eqnarray}
where $\alpha\leq\gamma$ is the \emph{exploration} hyperparameter, $\beta\in[0,1]$ is the \emph{regulation} or \emph{exploitation} parameter and $\ell_{expl}(x,u)$ is the info-gain from the model, similar to \citep{hafner_reliable_2018}:
\begin{equation}
    \ell_{expl}(x,u) = \sum_{i=1,\dots, N_x}\frac{\left(\Sigma_{ii}\left({x}, u, p, \theta\right)\right)^2}{N_x\ \sigma_{y_{ii}}^2}.  
\end{equation}

\paragraph{Probabilistic Safety.} We study the feasibility and stability of the proposed scheme, following the framework of \cite{rawlings_mayne_paper, rawlingsMPC}. In particular, if the MPC (\ref{eq:MPC_hard}) is \emph{always feasible}, and the terminal cost and terminal set satisfy (\ref{eq:stochastic}) with probability $1$, then the MPC (\ref{eq:MPC_hard}) enjoys some \emph{intrinsic robustness} properties. In other words, we should be able to control the physical system and come back to a neighborhood of the initial equilibrium point for any state in $\mathbb{X}_s$, the size of this neighborhood depending on the model accuracy. We assume a perfect solver is used and that the relaxed problems enforce the constraints exactly for a given $\gamma$.      

For the exploration MPC to be \emph{safe}, we wish to be able to find a $u^\star(t)$ satisfying the terminal constraint: $$V(\hat{x}^{+}(t))\leq l_s,\ \forall t,$$ starting from the stochastic system: $$\hat{x}^{+}(t)= x(t) + \left(\mu(x(t),u^\star(t),\theta_\mu)+w(t)\right)dt,\quad w(t) \sim\ \mathcal{N}\left(0,\Sigma(x(t),u^\star(t),\theta_\Sigma)\right).$$ We aim at a probabilistic result. First, recall that we truncate the distribution of, $w$, to a high confidence level z-score, $\bar{\sigma}$. 
Once again, we switch to a set-valued uncertainty representation as it is most convenient and provide a result that depends on the z-score $\bar{\sigma}$. Assume known the probability of our model to be able to perform one step predictions, given the model $\mathcal{M}$, such that the real state increments are within the given confidence interval, $\bar{\sigma}$, and define it as: $P(x(t+1) \in \mathcal{M}(x(t))|\mathcal{M},\bar{\sigma})=1-\epsilon_\mathcal{M}(\bar{\sigma})$.  This probability can be estimated and improved using cross-validation, for instance by fine-tuning  $\sigma_w$. It can also be increased with  $\bar{\sigma}$ after the model training. This can however make the control search  more challenging. Finally, since we use probabilistic verification, from (\ref{eq:stochastic}) we have a probability of the terminal set to be invariant for the model with truncated distributions: $P\left(\mathcal{R}(\mathbb{X}_s)\subseteq \mathbb{X}_s|\mathcal{M},\ \bar{\sigma}\right)=P\left(\mathcal{R}(\mathbb{X}_s)\subseteq \mathbb{X}_s\right)=1-\epsilon_p$, where $\mathcal{R}$ is the one-step reachability set operator \citep{Kerrigan:2000} computed using the model in closed loop with $K$. Note that this probability is determined by the number of verification samples \citep{bobiti_samplingdriven_nodate}. Safety of the next state is determined by:
\begin{theorem}
Given $x(t)\in\mathbb{X}_s$, the probability of (\ref{eq:MPC_hard}-\ref{eq:MPCexploration}) to be feasible (safe) at the next time step is: 
    \begin{align}
       P\left(x(t+1)\in\mathbb{X}_s|\mathcal{M},\ \bar{\sigma},\  x(t)\in\mathbb{X}_s\right) = \\ P(x(t+1) \in \mathcal{M}(x(t))|\mathcal{M},\ \bar{\sigma})P(\mathcal{R}(\mathbb{X}_s)\subseteq \mathbb{X}_s)  =\\ (1-\epsilon_\mathcal{M}(\bar{\sigma}))(1-\epsilon_p). 
    \end{align}
\end{theorem} 
It must be noticed that, whilst $P\left(\mathcal{R}(\mathbb{X}_s)\subseteq \mathbb{X}_s\right)$ is constant, the size of $\mathbb{X}_s$ will generally decrease for increasing $\bar{\sigma}$ as well as $\sigma_w$. The probability of any state to lead to safety in the next step is given by: 
\begin{theorem}
Given $x(t)$, the probability of (\ref{eq:MPC_hard}-\ref{eq:MPCexploration}) to be feasible (safe) at the next step is: 
    \begin{align}
       P_\text{SAFE}(\mathcal{M},\ \bar{\sigma}) = P(x(t) \in \mathcal{C}(\mathbb{X}_s))  P\left(x(t+1)\in\mathbb{X}_s|\mathcal{M},\ \bar{\sigma}, x(t)\in\mathbb{X}_s\right) \geq \\ P(x(t) \in \mathbb{X}_s) (1-\epsilon_\mathcal{M}(\bar{\sigma}))(1-\epsilon_p),
    \end{align}
\end{theorem} 
where $\mathcal{C}$ denotes the one-step robust controllable set for the model \citep{Kerrigan:2000}. The size of the safe set is a key factor for a safe system. This depends also on the architecture of $V$ and $K$ as well as on the stage cost matrices $Q$ and $R$. A stage cost is not explicitly needed for the proposed approach, however, $Q$ can be beneficial in terms of additional robustness and $R$ serves as a regularisation for $K$. 

\section{Network architecture for inverted pendulum}
\paragraph{Forward Model} Recall the NPC-BRNN definition:
\begin{align} \label{eq:forward_rev}
      \hat{x}(t+1) = \hat{x}(t) + dt\ d\hat{x}(t),& \quad 
      d\hat{x}(t) = \mu\left(\hat{x}(t), u(t); \theta_\mu\right) + \hat{w}(t), \\
      \hat{w}(t)  \sim q(\hat{x}(t), u(t);\theta_\Sigma), &\quad q(\hat{x}(t), u(t);\theta_\Sigma) = \mathcal{N}\left(0,\ \Sigma\left(\hat{x}(t), u(t); \theta_\Sigma\right)\right), \\
      \hat{y}(t) \sim \mathcal{N}\left(\hat{x}(t), \sigma_y^2\right), \label{eq:measure} & \quad 
      \hat{x}(0) \sim \mathcal{N}(x(0), \sigma_y^2).
  \end{align}
Partition the state as $x = \left[
     x_1 \ 
     x_2\right]^T$, 
where the former represents the angular position and the latter the velocity. They are normalised, respectively, to a range of $\pm\pi$ and $\pm2\pi$.  
We consider a $\mu$ of the form:
\begin{align} 
\mu(x,u) = \left[\begin{array}{cc}
    2\cdot x_2 \\
     \text{f}_\mu(x,u) 
\end{array}\right]
  \end{align}
where $\text{f}_\mu(x,u)$ is a three-layer feed-forward neural network with $64$ hidden units and $\tanh$ activations in the two hidden layers. The final layer is linear. The first layer of $\text{f}_\mu$ is shared with the standard deviation network, $\Sigma(x,u)$, where it is then followed by one further hidden layer of $64$ units before the final sigmoid layer. The parameter $\sigma_w$ is set to $0.01$. The noise standard deviation, $\sigma_y$ is passed through a softplus layer in order to maintain it positive and was initialised at $0.01$ by inverting the softplus. We used $1000$ epochs for training, with a learning rate of 1E-4 and an horizon of $10$. The sequences where all of $1000$ samples, the number of sequences was increased by increments of $10$ and the batch size adjusted to have sequences of length $10$. The target loss was initilisized as $-2.94$. 

We point out that this architecture is quite general and has been positively tested on other applications, for instance a double pendulum or a joint-space robot model, with states partitioned accordingly. 

\paragraph{Lyapunov function} The Lyapunov net consists of three fully-connected hidden layers with $64$ units and $\tanh$ activations which are then followed by a final linear layer with $200$ outputs. These are then reshaped into a matrix, $V_{net}$, of size $100\times 2$ and is evaluated as:
\begin{equation}\label{eq:lyap1_rev}
  V(x) = x^T\text{sofplus}(\alpha) \left(\epsilon I + V_{net}(x)^T V_{net}(x)\right)x + \psi(x),
  \end{equation}
where $\epsilon>0$ is a hyperparameter and $\alpha$ is a trainable scaling parameter which is passed through a softplus. The  introduction of $\alpha$ noticeably improved results. The prior function $\phi$ was set to keep $|x_1|\leq 0.3$. We used $61$ outer epochs for training and $10$ inner epochs for the updates of $V$ and $K$, with a with learning rate 1E-3. We used a uniform grid of $10k$ initial stastes as a single batch.   
\paragraph{Exploration MPC}
For demonstrative purposes we solved the Safe-MPC using Adam with $3000$ epochs for the minimisation step and SGD with $100$ epochs for the maximisation step. The outer loop used $3$ iterations. The learning rates were set to, respectively, $0.1$ and 1E-4. The exploration factor, $\alpha$, was set to $100$ as well as the soft constraint factor, $\gamma$. The exploitation factor, $\beta$, was set to $1$. 

\section{Considerations on model refinement and policies}
Using neural networks present several advantages over other popular inference models: for instance, their scalability to high dimensional and to large amount of data, the ease of including physics-based priors and structure in the architecture and the possibility to learn over long sequences.  At the same time, NNs require more data than other methods and no offer no formal guarantees. For the guarantees, we have considered a-posteriori probabilistic verification. For the larger amount of data, we have assumed that an initial controller exists (this is often the case) that can be used to safely collect as much data as we need. 

\paragraph{Model refinement.} A substantial difficulty was encountered while trying to incrementally improve the results of the neural network with increasing amount data. In particular, as more data is collected, larger or more batches must be used. This implies  that either the gradient computation or the number of backward passes performed per epoch is different from the previous model training. Consequentely, if a model is retrained entirely from scratch, then the final loss and the network parameters can be substantially different from the ones obtained in the previous trial. This might result in having a larger uncertainty than before in certain regions of the state space. If this is the case, then stabilising the model can become more difficult and the resulting safe set might be smaller. We have observed this pathology initially and have mitigated it by employing these particular steps: first, we use a sigmoid layer to effectively limit the maximum uncertainty to a known hyperparameter; second, we added a consistency loss that encourages the new model to have uncertainty smaller than the previous one over the (new) training set; third, we used rejection sampling for the background based on the uncertainty of the previous model, so that the NCP does not penalise \emph{previously known} datapoints; finally, we stop the training loop as soon as the final loss of the previous model is exceeded. These ingredients have proven successful in reducing this pathology and, together with having training data with increasing variance, have provided that the uncertainty and safe set improve up to $90k$ datapoints. After that, however, adding further datapoints has not improved the size of the safe set which has not reached its maximial possible size. We believe that closing the loop with exploration could improve on this result but are also going to investigate further alternatives.  

Noticeably, \cite{Gal2016Improving} remarked that improving their BNN model was not simple. They tried for instance to use a forgetting factor which was not successful and concluded that their best solution was to save only a fixed number of most recent trials. We believe this could not be sufficient for safe learning as the uncertain space needs to be explored. Future work will further address this topic, for instance, by   retraining only part of the network, or possibly by exploring combinations of our approach with the ensemble approach used in \cite{max}. Initial trials of the former seemed encouraging for deterministic models.  

\paragraph{Training NNs as robust controllers.} In this paper, we have used a neural network policy for the computation of the safe controller. This choice was made fundamentally to compare the results with an LQR, which can successfully solve the example. Training policies with backpropagation is not an easy task in general. For more complex scenarios, we envisage two possible solutions: the first is to use evolutionary strategies \citep{stanley2002evolving, salimans2017es} or other global optimisation methods to train the policy; the second is to use a robust MPC instead of a policy. Initial trials of the former seemed encouraging. The latter would result in a change of Algorithm \ref{alg:alternateDescent}, where the $K$ would be not learned but just evaluated point-wise through an MPC solver. Future work is going to investigate these alternatives.   

\section{Related work}

\paragraph{Robust and Stochastic MPC} Robust MPC can be formulated using several methods, for instance: min-max optimisation \citep{Bemporad_minmax, Kerrigan2004, Raimondo2009}, tube MPC \citep{rawlingsMPC, Rakovic2012} or constraints restriction \citep{richards_a._g._robust_2004} provide robust recursive feasibility given a known bounded uncertainty set. 
In tube MPC as well as in constraints restriction, the nominal cost is optimized while the constraints are restricted according to the uncertainty set estimate. This method can be more conservative but it does not require the maximization step. For non-linear systems, computing the required control invariant sets is generally challenging. Stochastic MPC approaches the control problem in a probabilistic way, either using expected or probabilistic constraints. For a broad review of MPC methods and theory one can refer to \cite{Maciejowski_book, Camacho2007, rawlingsMPC,  Cannon_book, Gallieri2016, Borrelli_book, Rakovic2019}. 

\paragraph{Adaptive MPC} \cite{lorenzen_cannon} presented an approach based on tube MPC for linear parameter varying systems using set membership estimation. In particular, the constraints and model parameter set estimates are updated in order to guarantee recursive feasibility. \cite{pozzoli_tustin_2019} used the Unscented Kalman Filter (UKF) to adapt online the last layer of a novel RNN architecture, the Tustin Net (TN), which was then used to successfully control a double pendulum though MPC. TN is a deterministic RNN which is related to the architecture used in this paper. A comparison of different network architectures, estimation and adaptation heuristics for neural MPC can be found, for instance, in \cite{pozzoli}. 

\paragraph{Stability certification}
\cite{bobiti_sampling-based_2016} proposed a grid-based deterministic verification method which relies on local Lipschitz bounds of the system dynamics. This approach requires knowledge of the model equations and it was extended to black-box simulations  \citep{bobiti_samplingdriven_nodate} using a probabilistic approach. We extend this framework by means of a check for ultimate boundedness and propose to use it with Bayesian models through uncertainty sampling.  

 \paragraph{MPC for Reinforcement Learning}
 \cite{Williams2017} presented an information-theoretical framework to solve a non-linear MPC in real-time using a neural network model and performed model-based RL on a race car scale-model with non-convex constraints. \cite{gros_towards_2019} used policy gradient methods to learn a classic robust MPC for linear systems.

\paragraph{Safe learning} \cite{Yang2015b,Yang2015} looked, respectively, at using GPs for risk sensitive and fault-tolerant MPC. \cite{vinogradska_gaussian_nodate} proposed a quadrature method for computing invariant sets, stabilising and unrolling GP models for use in RL. \cite{berkenkamp_safe_2017} used the deterministic verification method of \citep{bobiti_sampling-based_2016} on GP models and embedded it into an RL framework using approximate dynamic programming. The resulting policy has a high probability of safety. \cite{Akametalu2014} studied the reachable sets of GP models and proposed an iterative procedure to refine the stabilisable (safe) set as more data is collected.  \cite{hewing_cautious_2017} reviewed uncertainty propagation methods and formulated a chance-constrained MPC for grey-box GP models. The approach was demonstrated on an autonomous racing example with non-linear constraints.  \cite{Limon2017} presented an approach to learn a non-linear robust model predictive controller based on worst case bounding functions and \:{H}older constant estimates from a non-parametric method. In particular, they use both the trajectories from an initial offline model for recursive feasibility as well as of an online refined model to compute the optimal loss.
\cite{koller_learning-based_2018} provide high probability guarantees of feasibility for a GP-based MPC with Gaussian kernels. This is done using a closed-form exact Taylor expansion that results in the solution of a generalised eigenvalue problem per each step of the prediction horizon. 
\cite{cheng_end--end_2019} complemented model-free RL methods (TRPO and DDPG) with a GP model based approach using a barrier function safety loss, the GP being refined online. \cite{chow_lyapunov-based_2018} developed a safe Q-learning variant for constrained Markov decision processes based on a state-action Lyapunov function. The Lyapunov function is shown to be equal to the value function for a safety constraint function, defined over a finite horizon. This is constructed by means of a linear programme. \cite{chow_lyapunov-based_2019} extended this approach to policy gradient methods for continuous control. Two projection strategies have been proposed to map the policy into the space of functions that satisfy the Lyapunov stability condition. \cite{Papini2018SafelyEP} proposed a policy gradient method for exploration with a statistical guarantee of increase of the value function.   \cite{wabersich_probabilistic_2019} formulated a probabilistically safe method to project the action resulting from a model-free RL algorithm into a safe manifold. Their algorithm is based on results from chance constrained tube-MPC and makes use of a linear surrogate model. \cite{thananjeyan_safety_2019} approximated the model uncertainty using an ensemble of recurrent neural networks. Safety was approached by constraining the ensemble to be close to a set of successful demonstrations, for which a non-parametric distribution is trained. Thus, a stochastic constrained MPC is approximated by using a set of the ensemble models trajectories. The model rollouts are entirely independent. Under several assumptions the authors  proved the system safety. These assumptions can be  rarely met in practise, however, the authors demonstrated that the approach works practically on the control of a manipulator in non-convex constrained spaces with a low ensemble size.

\paragraph{Learning Lyapunov functions}
\cite{verdier_formal_2017} used genetic programming to learn a polynomial control Lyapunov function for automatic synthesis of a continuous-time switching controller.
\cite{richards_lyapunov_2018} proposed an architecture and a learning method to obtain a Lyapunov neural network from labelled sequences of state-action pairs. Our Lyapunov loss function is inspired by \cite{richards_lyapunov_2018}  but does not make use of labels nor of sequences longer than one step. These approaches were all demonstrated on an inverted pendulum simulation. \cite{taylor_episodic_2019} developed an episodic learning method to iteratively refine the derivative of a continuous-time Lyapunov function and improve an  existing controller solving a QP. Their approach exploits a factorisation of the Lyapunov function derivatives based on feedback linearisation of robotic system models. They test the approach on a segway simulation.  

\paragraph{Planning and value functions}
POLO \citep{lowrey_plan_2018} consists of a combination of online planning (MPC) and offline value function learning. The value function is then used as the terminal cost for the MPC, mimicking an infinite horizon. The result is that, as the value function estimation improves, one can, in theory, shorten the planning horizon and have a near-optimal solution. The authors demonstrated the approach using exact simulation models. This work is related to SiMBL, with the difference that our terminal cost is a Lyapunov function that can be used to certify safety. 

\paragraph{Uncertain models for RL}
PILCO \citep{DeisenrothRT2011, deisenroth_gaussian_2015} used of GP models for model-based RL in an MPC framework that trades-off exploration and exploitation. \cite{Frigola2014VariationalGP} formulated a variational GP state-space model for time series. \cite{Gal2016Improving} showed that variational NNs with dropout can significantly outperform GP models, when used within PILCO, both in terms of performance as well as computation and scalability.  
\cite{chua_deep_2018} proposed the use of ensemble RNN models and an MPC-like strategy to distinguish between noise and model uncertainty. They plan over a finite horizon with each model and optimise the action using a cross-entropy method. \cite{kurutach_model-ensemble_2018} used a similar forward model for trust-region policy optimisation and showed significant improvement in sample efficiency with respect to both single model (no uncertainty) as well as model-free methods. MAX \citep{max} used a similar ensemble of RNN models for efficient exploration, significantly outperforming baselines in terms of sample efficiency on a set of discrete and continuous control tasks.  
\cite{depeweg_learning_2016} trained a Bayesian neural network using the $\alpha$-divergence and demonstrated that this can outperform both variational networks, MLP and GP models when used for stochastic policy search over a gas turbine example with partial observability and bi-modal distributions. \cite{hafner_learning_2018} proposed to use an RNN with both deterministic and stochastic transition 
components together with a multi-step variational inference objective. Their framework predicts rewards directly from pixels. This differs from our approach as we don't have deterministic states and consider only full state information. \cite{Carron2019} tested the use of a nested control scheme based on an internal feedback linearisation and an external chance-constrained offset free MPC. The MPC is based on nominal linear models and uses both a sparse GP disturbance model as well as a piece-wise constant offset which is estimted online via the Extended Kalman Fitler (EKF). The GP uncertainty is propagated through a first order Taylor expansion. The approach was tested on a robotic arm.

\section*{Acknowledgements}
The authors are grateful to Christian Osendorfer, Boyan Beronov, Simone Pozzoli, Giorgio Giannone, Vojtech Micka, Sebastian East, David Alvarez, Timon Wili, Pierluca D'Oro, Wojciech Ja\'{s}kowski, Pranav Shyam, Mark Cannon and Andrea Carron for constructive discussions. We are also grateful to Felix Berkenkamp for the support given while experimenting with their safe learning tools. All of the code used for this paper was implemented from scratch by the authors using PyTorch. Finally, we thank everyone at NNAISENSE for contributing to a successful and inspiring R\&D environment.  

\end{document}